\documentclass[]{pasj01}
\draft

\begin{document} 
\Received{}
\Accepted{}

\title{Searching for Moving Objects in HSC-SSP: Pipeline and Preliminary Results}


\author{Ying-Tung \textsc{Chen}\altaffilmark{1}}
\altaffiltext{1}{Institute of Astronomy and Astrophysics, Academia Sinica, P. O. Box 23-141, Taipei 106, Taiwan}
\email{ytchen@asiaa.sinica.edu.tw}

\author{Hsing-Wen \textsc{Lin}\altaffilmark{2}}%
\altaffiltext{2}{Institute of Astronomy, National Central University, 32001, Taiwan}

\author{Mike \textsc{Alexandersen}\altaffilmark{1}}
\author{Matthew J. \textsc{Lehner}\altaffilmark{1,3,4}}
\altaffiltext{3}{Department of Physics and Astronomy, University of Pennsylvania, 209 S. 33rd St., Philadelphia, PA 19125,
USA}
\altaffiltext{4}{Harvard-Smithsonian Center for Astrophysics, 60 Garden St., Cambridge, MA 02138, USA}
\author{Shiang-Yu \textsc{Wang}\altaffilmark{1}}

\author{Jen-Hung \textsc{Wang}\altaffilmark{1}}
\author{Fumi \textsc{Yoshida}\altaffilmark{5,6}}
\altaffiltext{5}{Planetary Exploration Research Center, Chiba Institute of Technology, 2-17-1 Tsudanuma, Narashino, Chiba
275-0016, Japan}
\altaffiltext{6}{Department of Planetology, Graduate School of Science, Kobe University, Kobe, 657-8501, Japan}

\author{Yutaka \textsc{Komiyama}\altaffilmark{7}}
\altaffiltext{7}{National Astronomical Observatory of Japan, 2-21-1 Osawa, Mitaka, Tokyo 181-8588, Japan }

\author{Satoshi \textsc{Miyazaki}\altaffilmark{7,8}}
\altaffiltext{8}{SOKENDAI(The Graduate University for Advanced Studies), Mitaka,
  Tokyo, 181-8588, Japan}



\maketitle

\begin{abstract}

The Hyper Suprime-Cam Subaru Strategic Program (HSC-SSP) is currently the deepest wide-field survey in progress. The 8.2~m
aperture of Subaru telescope is very powerful in detecting faint/small moving objects, including near-Earth objects,
asteroids, centaurs and Tran-Neptunian objects (TNOs). However, the cadence and dithering pattern of the HSC-SSP are not
designed for detecting moving objects, making it difficult to do so systematically.
In this paper, we introduce a new pipeline for detecting moving objects (specifically TNOs) in a non-dedicated survey. 
The HSC-SSP catalogs are re-arranged into the {\tt HEALPix} architecture. Then, the stationary detections and false positive
are removed with a machine learning algorithm to produce a list of moving object candidates. An orbit linking algorithm and
visual inspections are executed to generate the final list of detected TNOs.
The preliminary results of a search for TNOs using this new pipeline on data from the first HSC-SSP data release (Mar 2014 to
Nov 2015) are also presented.
\end{abstract}

\newpage

\section{Introduction}
Minor planets are the proxy to understand the history and evolution of the Solar System. 
In the past two decades, several surveys have used different facilities to discover near-Earth asteroids (NEAs), main belt
asteroids (MBAs), Centaurs and trans-Neptunian Objects (TNOs). 
For the nearby objects (NEAs, MBAs), small telescopes with quick response times are very efficient at detecting them and are
thus useful to help understand the population of Earth's neighborhood. 
For more distant objects (Centaurs, TNOs), large telescopes are used to gather more photons from objects with less reflecting
light \citep{ell05, kai10, pet11, ale16, ban16}.
Due to the difficulty of obtaining accurate data on small objects at such large distances, the physical properties of only the
brightest, largest objects have been well studied.
However, the fainter, smaller objects can provide important clues to understand the collisional and accretion processes in the
evolution of the Solar System. Recent spacecraft missions, including the Rosetta mission to the comet 67P/Churyumov–Gerasimenko \citep{for15}
and the Hayabusa mission to 25143 Itokawa \citep{fuj06} reveal that our understanding of small Solar System objects is quite limited.

The Hyper Suprime-Cam Subaru Strategic Program (HSC-SSP) is a large program on the 8.2~m Subaru telescope, using the large
1.7~deg$^2$. field-of-view (FoV) camera, HSC, to survey around 1400 deg$^2$ of the sky. 
The uniqueness of this facility makes HSC-SSP the deepest wide-field survey currently in operation. 
Having started in 2014, HSC-SSP will use 300 Subaru telescope nights over 6 years (2014-2019).

The cadence of a wide-field survey is very crucial to efficiently detect moving objects. 
For a dedicated survey of TNOs, the survey cadence usually requires a pair or a triplet of exposures with one to a few hours
separation and the same pointing to discover the objects. 
The Deep Ecliptic Survey (DES) was one of the first wide-field surveys focused on the outer Solar System objects
\citep{ell05}. 
After that, the Canada-France Ecliptic Plane Survey (CFEPS) was the first dedicated survey with a full characterization
including orbital parameter determination and detection efficiencies \citep{pet04, kav09, pet11}.
CFEPS used a set of three exposures ($\sim$1 hr cadence) to resolve the motions of TNOs.
Following CFEPS, the Panoramic Survey Telescope and Rapid Response System 1 (Pan-STARRS 1, or PS1) project searched for Solar
System objects in a 3$\pi$ area of the sky.
The major goals of PS1 were to detect NEAs \citep{wai14} and TNOs \citep{chen16, lin16} with a sub-hour cadence.
In the above projects, a specific pipeline was designed to operate with the preferred cadence.
The PS1 moving objects pipeline, the Moving Object Processing System (MOPS, \cite{den13}), was applied to detect asteroids and
NEAs, while the PS1 Outer Solar System (OSS) team developed a different pipeline to detect TNOs \citep{hol15}.

The major goal of HSC-SSP is to address outstanding astrophysical questions such as the nature of dark matter and dark energy,
the cosmic re-ionization, and galaxy evolution over cosmic timescales. 
In this survey, a wide-dithering pattern with sub-hour cadence under similar weather condition is often applied to fill up the
gaps between CCDs of HSC in order to obtain complete coverage of a targeted field. 
Such a survey strategy makes it extremely difficult to detect moving objects using the conventional techniques used by the TNO
surveys described above. Also, the high number density of the detections in HSC image results
in unreasonable requirements for CPU time if the pipeline uses the extensively adopted KD-Tree \citep{ben75, kub05} algorithm.
However, we would like to utilize the data set from the HSC survey to search for small TNOs. Finding more faint TNOs is
essential to determine the faint end of the TNO size distribution, and the wide field and depth of HSC-SSP has a unique
potential to find a large number of such objects.

In order to search for TNOs in this data set, we thus needed to develop a new moving object pipeline, HSC-MOP, to detect Solar
System objects in the HSC-SSP data. The pipeline is capable of detecting the moving objects in a wide-field survey with a
non-uniform dithering cadence like that of the HSC-SSP. This pipeline has the ability to detect moving objects in images with
different time separations and different limiting magnitudes.

The rest of this paper is organized as follows. In section~\ref{sec:data}, we introduce the first data release of the HSC-SSP,
which includes observations taken from 2014 to 2015 \citep{aih17}. In section~\ref{sec:pipe}, we describe our moving object
pipeline. In section~\ref{sec:result}, we present the preliminary results of a search for TNOs using our pipeline on the first
HSC-SSP data release. Finally, in section~\ref{sec:con}, we conclude the paper with a summary of our results.


\section{Data status} \label{sec:data}
The HSC-SSP survey is executed with HSC which consists of 116 2048 x 4176 pixel CCDs, with a pixel scale of $0.168\arcsec$.
\citep{miy12}. 104 of the CCDs are for science.
With a 1.76 deg$^2$ FoV, HSC-SSP will  survey $\sim$1400~deg$^2$ using Sloan-like $g$, $r$, $i$, $z$, and $y$ broadband
filters. The first data set was released on 27 February 2017\footnote{http://hsc.mtk.nao.ac.jp/ssp/} \citep{aih17}.

There are three layers of the HSC-SSP survey including WIDE, DEEP, and UDEEP fields, each with different coverage areas and
limiting magnitudes. 
We only process the WIDE layer of data in this study to maximize the FoV for the search of TNOs. 
Note that the first data release of HSC-SSP only include the sky regions covered in all the five filters, which is different
from what we analysed in this study -- we did not require the fields to be imaged in $z$ or $y$ bands.
The HSC-SSP observations are acquired in patches, each patch covering a sky area formed by a dithered layout of the
1.7~deg$^2$ HSC FoV, such that the gaps between CCD chips could be covered and the survey area would have the required
exposure depth for the primary science goals.

Because the weather conditions and schedule arrangements vary in each run, the HSC-SSP has a flexible schedule for the
sequence of observations. As a result, the patches observed with each filter in a single dark run have different sizes.
In this study, we analyzed the images with deepest limiting magnitudes i.e. $\sim$3 minute exposures taken in HSC-$g$,
HSC-$r$, HSC-$i$, HSC-$r2$, and HSC-$i2$ filters. 
The calibration images using shorter exposure time ($\sim30$ second) are not included in this study.
The limiting magnitude of each HSC-$r$ exposure is around 25.0~to 25.5~mag. 
Note that the original $r$- and $i$-band filters of HSC (“HSC-$r$” and “HSC-$i$” ) were replaced with new ones ( “HSC-$r2$”
and “HSC-$i2$” ) in June 2016 and Nov 2015, respectively. The transmission curve for the old filters and the new ones are
similar, but the new HSC-r2 and HSC-i2 filters have more uniform transparency than HSC-r and HSC-i across the effective area.
There is no data using new filters in this study.

\begin{figure}
\includegraphics[width = 1.\textwidth]{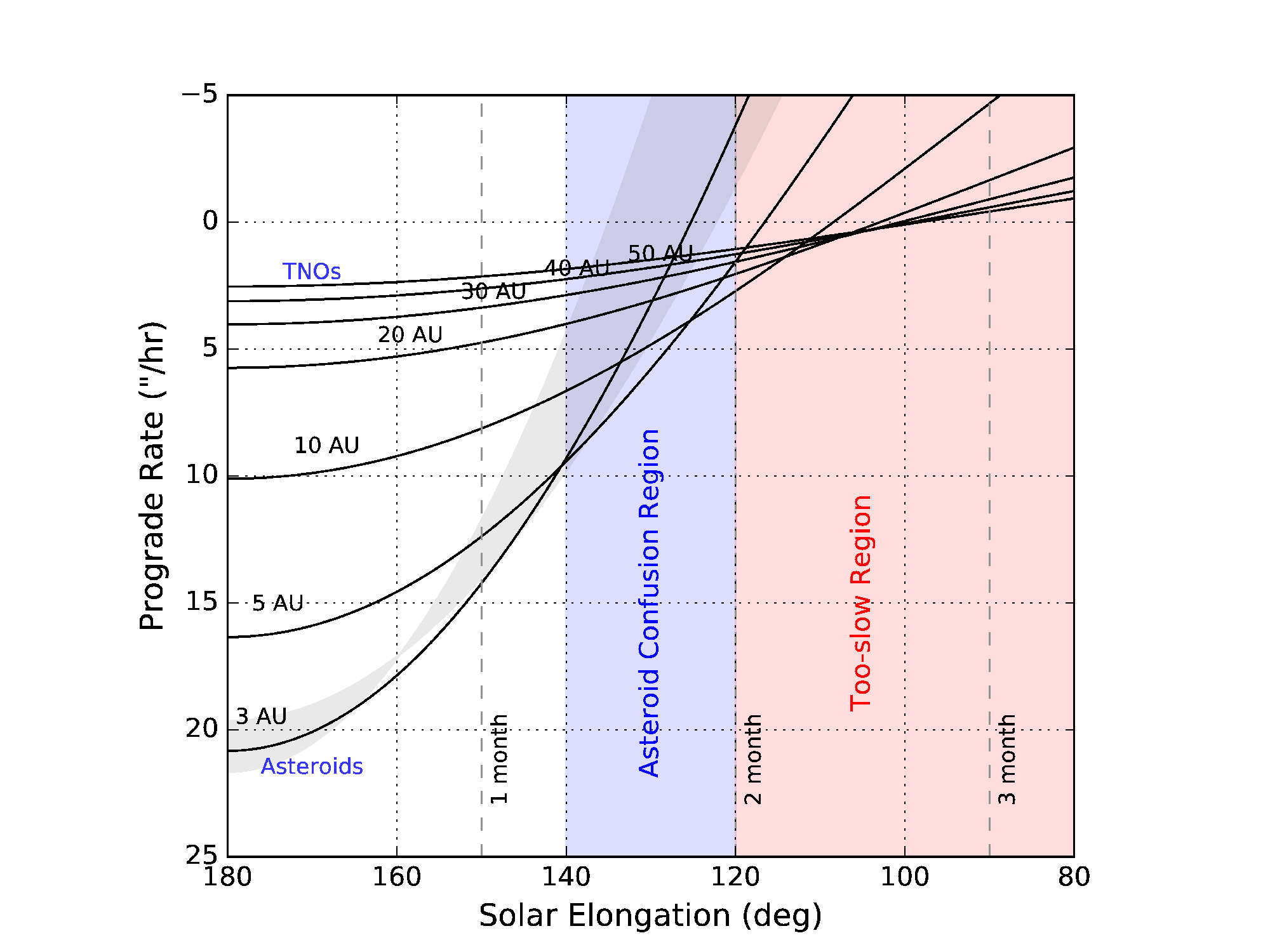} 
\caption{Rate of motion versus solar elongation angle. The vertical dashed lines show the solar elongation angles
corresponding to the time to opposition. The blue rectangle indicates that the TNOs and asteroids have similar rates
of motion. The gray region indicates the rates of motion for main belt asteroids.}\label{fig:ac}
\end{figure}

\begin{figure}
\includegraphics[width = 1.\textwidth]{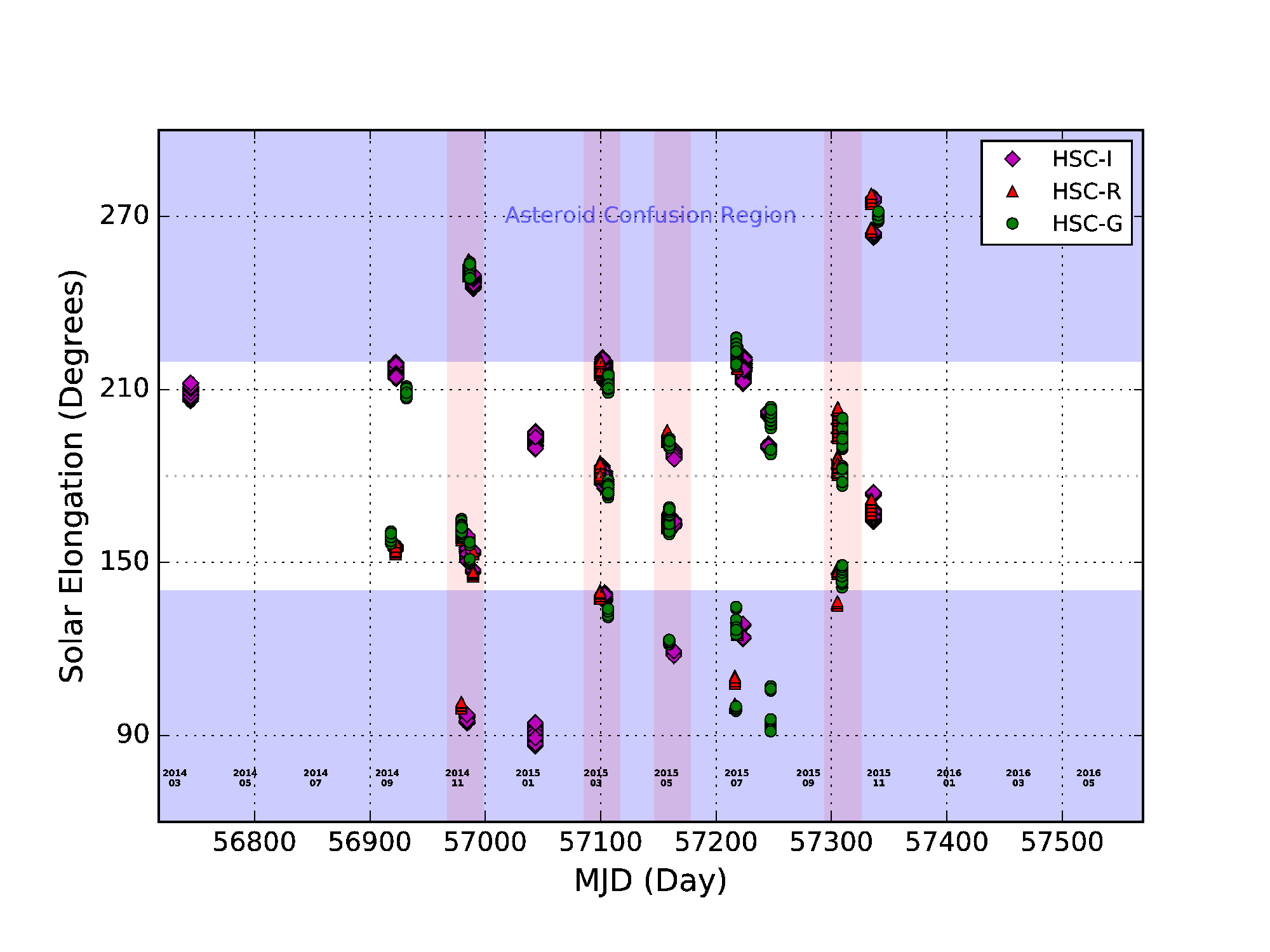} 
\caption{The solar elongations of HSC fields for images taken 2014--2015. The red regions indicate the dark runs with enough
exposures for data analysis, and the blue regions are excluded from this analysis due to the large solar
elongations.}\label{fig:se}
\end{figure}

\begin{figure}
\includegraphics[width = 1.\textwidth]{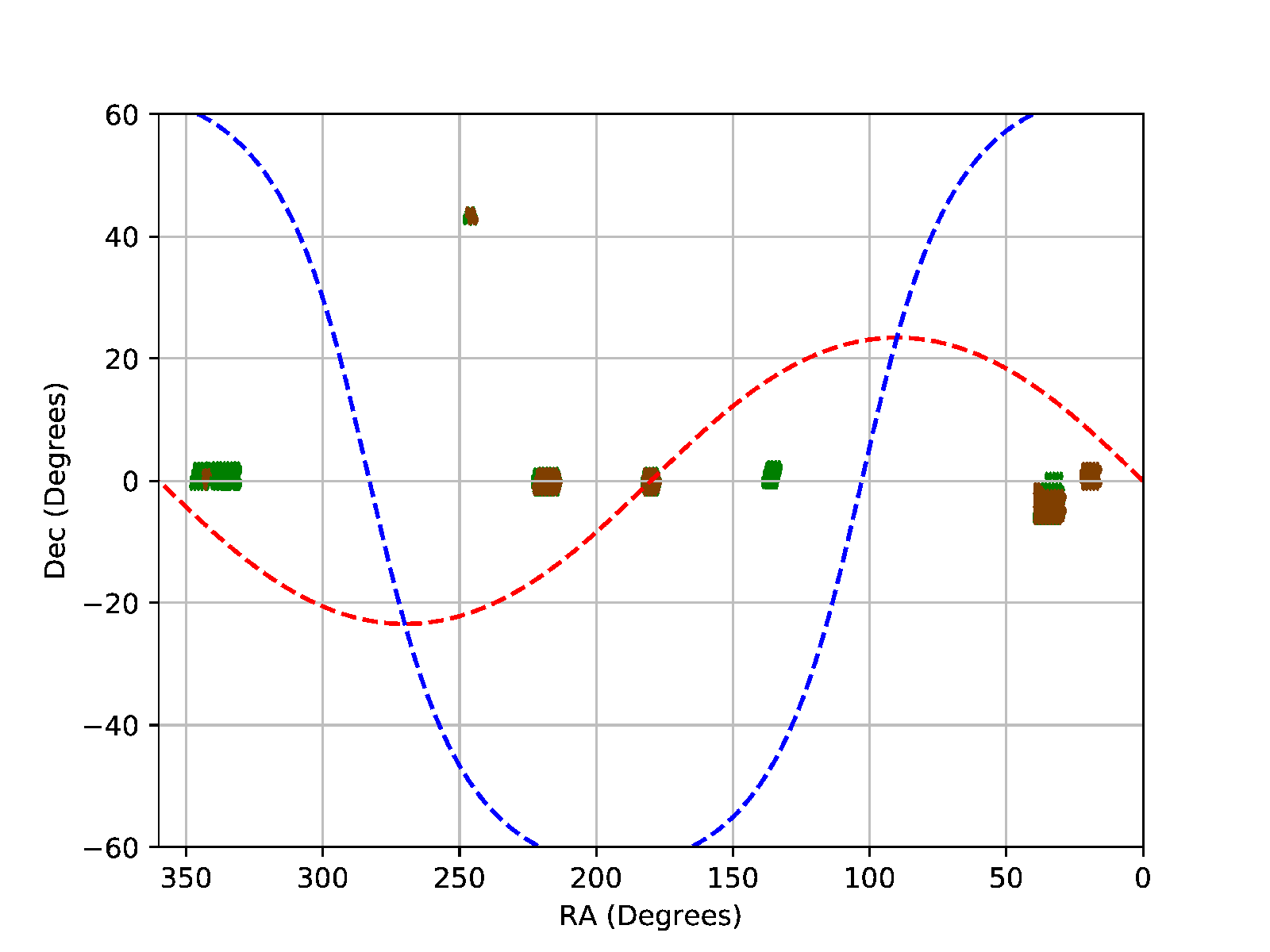} 
\caption{The HSC-SSP survey fields used in this analysis. Only $g$, $r$, and $i$ exposures within $\pm$ 40 degree to
opposition in the first data release were considered. The ecliptic plane and galactic plane are showed with dashed lines in
red and blue. The green regions indicate the total area of all FoVs using g, r, i filters. The brown regions show the areas
used in our moving object search. Note the sky regions is different from the first data release of HSC, which only include the
area covered in all the five filters.}\label{fig:area}
\end{figure}

The solar elongation is a crucial celestial parameter for observing Solar System objects. Due to phase effects, objects are
fainter as their locations farther from opposition. Furthermore, 
for objects observed at opposition ($\sim180\arcdeg$ solar elongation angle), the parallax from the motion of the Earth
dominates the apparent motion of objects across the sky (except for NEAs), such that the rate of apparent motion is strongly
dependent on the distance of the objects. 
The actual motion of the objects only provides substantial contributions to the apparent motion once the solar elongation is
$<140\arcdeg$ or $>220\arcdeg$ ($>40\arcdeg$ from opposition). 
The rates of motion of asteroids and TNOs are thus occasionally identical when the observations are far from opposition,
causing confusion (see Figure~\ref{fig:ac}).
Because the HSC-SSP is not a dedicated survey for Solar System science, many observations are acquired far from opposition. 
To avoid possible confusion of our detections, we limit our search to only data taken within $\pm40$ degrees from opposition
(see Figure~\ref{fig:se}). With this limit, $\sim221$ deg$^2$ out of the 333.5 deg$^2$ survey area taken from March 2014 to
November 2015 was used for this study (see Figure~\ref{fig:area} and Table~\ref{tab:first}). 


\section{Pipeline} \label{sec:pipe}
This section contains details about each step of our moving object detection pipeline - HSC-MOP. 
We begin with a quick overview, followed by detailed explanations of each step. 

\subsection{Overview}
\begin{figure}
\includegraphics[width = 1.\textwidth]{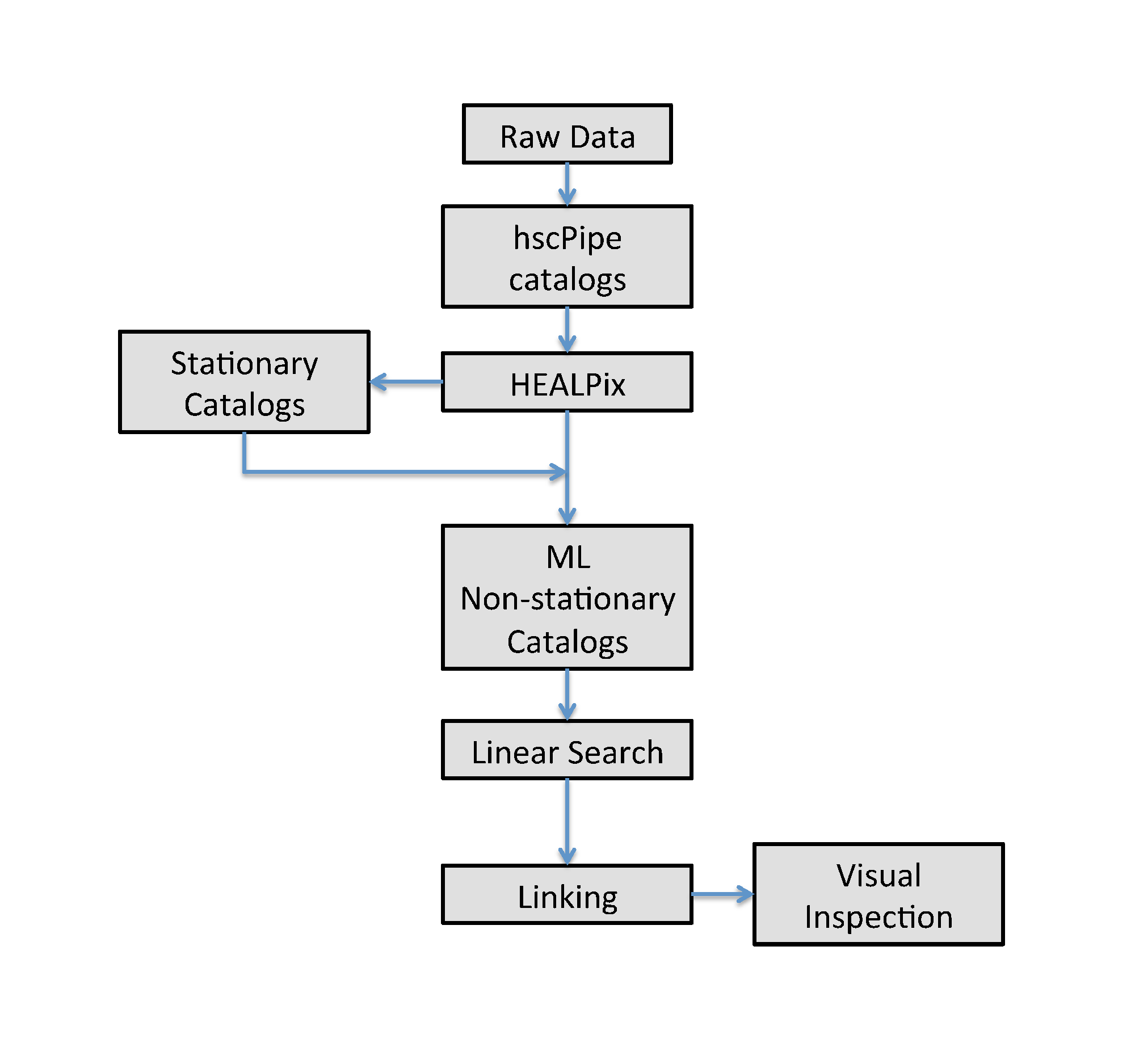} 
\caption{The flowchart of our pipeline.}\label{fig:fc}
\end{figure}

Our data process starts from the reduction of raw images with {\tt hscPipe}, the official HSC-SSP pipeline \citep{bos17}. 
A catalog of detected sources is created for each exposure generated from {\tt hscPipe}. Each field is then divided into
different sky regions using {\tt HEALPix} \citep{gor05}. The catalogs for each exposure which overlap each HEALPix region are
combined in to a master catalog. Objects which appear at the same place in multiple exposures are entered into a stationary 
catalog. A combination of machine learning algorithm and {\tt hscPipe} flags are then used to find candidate non-stationary
(that is, moving) sources.
After we build catalogs of non-stationary sources, a linear search algorithm is applied to identify possible moving objects. 
Such candidates would appear as non-stationary objects in different exposures with offset positions consistent with a moving 
object. These candidates are then verified with Initial Orbit Determination (IOD, see \cite{bk00}) to generate the list of
possible  candidates. Finally, all candidates are checked by visual inspection, using the Zooniverse platform. 
The Figure \ref{fig:fc} shows a flowchart of this pipeline.

\subsection{{\tt hscPipe}}
{\tt hscPipe} is the official pipeline to process data obtained from HSC, and is maintained by the HSC-SSP team. 
The standard outputs from {\tt hscPipe} include a fits table catalog and calibrated images (debiased, flat-fielded). 
{\tt hscPipe} uses the internal Pan-STARRS~1 catalog ($ps1\_pv2$) to calibrate astrometry and photometry
\citep{sch12, ton12, mag13}. 
The average astrometric scatter in an individual chip image is smaller than 30 milliarcsec. 
It should be noticed that the current version of {\tt hscPipe} still has some limitations on processing CCD chips
containing very bright stars.

\subsection{{\tt HEALPix}}
\begin{figure}
\includegraphics[width = 1.\textwidth]{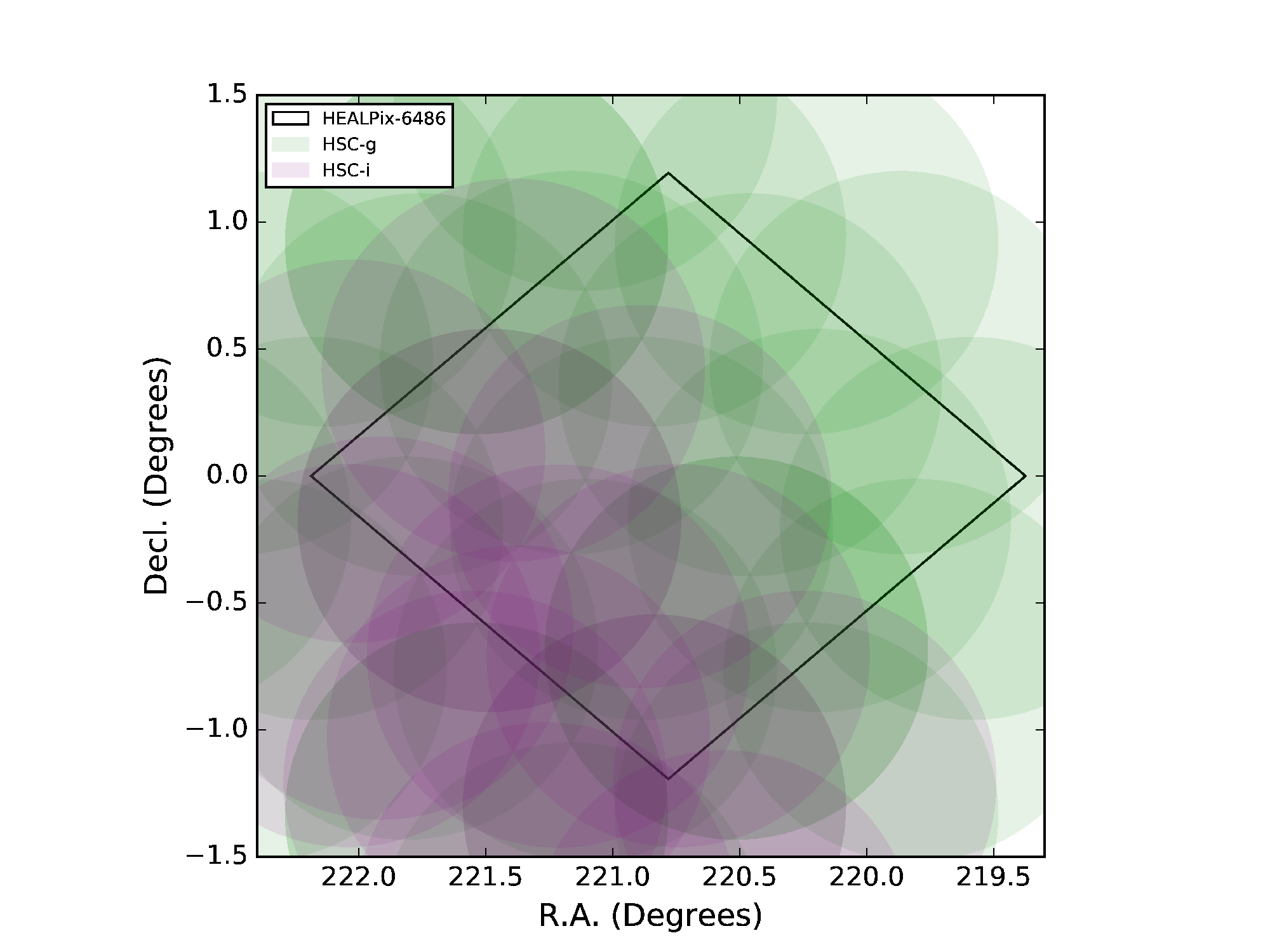}
\caption{A comparison of the HSC-SSP dithering pattern and the size of a {\tt HEALPix} region. The HSC-g (green) and HSC-i
(purple) observations are taken in the dark run of May 2015. The size of the {\tt HEALPix} region was made by the pixelation
with nside = 32.}\label{fig:hp}
\end{figure}

To handle the dithering observations, we adopted the Hierarchical Equal Area isoLatitude Pixelation of a sphere ({\tt
HEALPix}) \citep{gor05} to manage the catalogs acquired in a dark run.
The {\tt HEALPix} is an algorithm to produce a subdivision of the all sky, each pixel covers the same surface area as every
other pixel.
We adopt nside = 32 (12288 pixels for all sky) in the pipeline to build a set of reasonably sized sky regions so that the
pipeline can detect slower-moving objects within the same {\tt HEALPix}.
The mean size for each {\tt HEALPix} is 1.8323 deg (equals to 3.36 deg$^2$, see Figure \ref{fig:hp}) to match the HSC FoV.
Every detection catalog is separated and saved according to the corresponding {\tt HEALPix}.

\subsection{stationary catalog}
Most of the sources in the sky are stationary sources, i.e. background stars and galaxies. 
Detections of stationary sources should be at identical locations in different exposures, no matter when the images were
taken.
However, depending on the search criteria, moving objects might be cataloged as stationary objects. 
For example, objects beyond 50 au could be confused for stationary sources if the time-span between the images is not long 
enough for the object to move noticeably.
To avoid possible removal of the moving objects beyond the Kuiper Belt, we define a stationary object as (1) detected in at 
least two images within a radius of $0.5\arcsec$, and (2) with the separation between the image epochs longer than 20~minutes.
We use $0.5\arcsec$ instead of a larger radius after considering the HSC-SSP cadence, the false positive rate and the average 
seeing conditions of the data. 
Due to the cadence limitation of SSP, we could not choose a shorter maximum separation of the matched detection epochs. 
Under such conditions, only objects located at oppostion and further than 100 au might erroneously labeled as stationary objects. 
All exposures in each dark run are used to generate the stationary catalog, which is then applied in the next step which to 
remove the stationary detections and false positives.

\subsection{Machine learning to remove false positive}
After removing the stationary sources, there are still $\sim$0.5~million unclassified source detections in each {\tt HEALPix}
region. 
In addition to the real moving objects and transient events, the unclassified detections also include some
stars and galaxies with bad centroids and other possible artifacts.  
Therefore, processing all such detections without any further reductions will significantly impact the
performance and speed of the moving object detection pipeline. 
To conquer this problem, we adapted a machine learning (ML) based real-bogus system and {\tt hscPipe} flags to pick-up 
real astronomical detections and to reject the false positives. 
This system separates the real non-stationary sources and false positives by using 49 parameters generated from the {\tt
hscPipe}. These parameters mostly comprise the various photometric measurements (flux and error) and shape moments. 
Details of the 49 parameters are presented in \citet{lin17}.

Our real-bogus system can reach a true positive rate (tpr) of $\sim$96\% with a false positive rate (fpr) of $\sim$1\%, tpr 
$\sim$99\% with fpr $\sim$5\%, or tpr $\sim$99.5\% with fpr $\sim$15\%. We found that with the $1$\% fpr threshold, the system
can reduce the number of unclassified detections by half. 
If we relax the threshold to $15$\% fpr, it can still reduce the number of unclassified detections by one third. 
After several tests, we found that the pipeline could handle the remaining two thirds of the detections. 
We therefore set the threshold of fpr at $15$\% so that $99.5$\% of the real detections get passed, while $\sim$85\% of the 
original false detections get removed. 
For example, there are $\sim$0.5 million non-stationary detections in the {\tt HEALPix} for one SSP observation run without 
any false-positive cut. With the real-bogus system, $\sim$160,000 potentially false detections are rejected. The remaining 
sources are dominated by the faint real astronomical sources (stars, galaxies and etc). Since the detectability of very faint 
objects is highly sensitive to image quality (e.g., seeing and transparency), some of them cannot be removed by the stationary
selection criteria.
For details of the real-bogus system, we refer the reader to \citet{lin17}.

\subsection{grid linear search}
The next step of the pipeline is to search for matched detections of a candidate moving object in a single night. 
To maximize the number of candidates, we developed a new algorithm to detect possible moving objects, with the following 
steps. 

First, all non-stationary catalogs are merged into a single fits table, which includes the time, zero-point and image file 
path. The zero-point is used for calculating magnitude, while the image file path is for making postage stamp images to be 
used in the following step (see section~\ref{sec:zoo}).
We use the {\tt Python} module {\tt Numpy} to generate the catalogs. 

\begin{figure}
\includegraphics[width = 1.\textwidth]{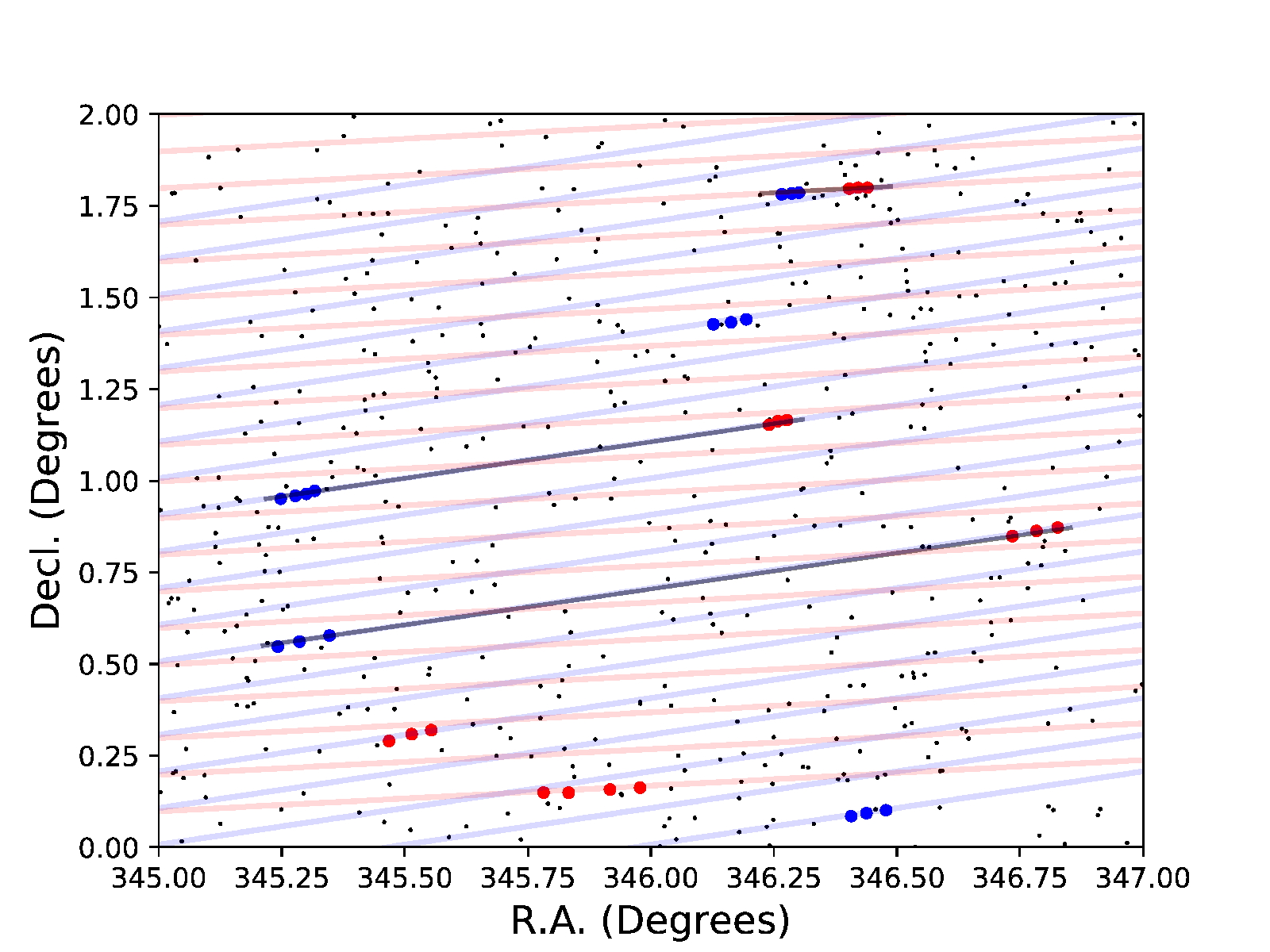} 
\caption{Illustration of the linear search. The sampling red and blue lines reveal the searches along different directions of
motion. The small black dots are objects in the non-stationary catalog, while the large blue and read points are examples of detected tracks in day 1 and day 2. The gray lines indicate the candidates with successful linkages}\label{fig:lin}
\end{figure}

Second, we use a simple Monte Carlo simulation to obtain the mean direction of motion of TNOs on the sky region at the time
of the dark run. Then the linear search scans the full area of a {\tt HEALPix} region within a range of $\pm30\arcdeg$ of the
resulting mean direction of motion.
Detections located within $2\arcsec$ of a line along a given direction of motion are grouped as a subset, which will be
verified if any reasonable group exists. This is demonstrated in Figure~\ref{fig:lin}.

With the grouped detections (sorted by RA), reasonable tracks (that is, groups of source detections spanning multiple images 
consistent with moving objects) are then identified using the following conditions: (1) correct 
time order along the line, (2) rates of motion $< 49\arcsec$/hr (ranging from asteroids to TNOs), (3) $\Delta$mag $<$ 2.0
across the different images, and (4) more than 3 detections. 
After the selection, the tracks of the candidates of moving objects are saved into a database.
\subsection{linking}
\label{sec:linking}
After we get a track of a candidate object in images taken over a single night, we link tracks from different nights that
might belong to the same object. 
The criteria are set by (1) similar rate and direction of motion, (2) source positions across the multiple
images consistent with a reasonable orbit. If the combination of two tracks satisfies the above criteria, the joined tracks
are then examined by IOD \citep{bk00}, which provides estimates of the orbital parameters, in particular semi-major axis $a$
and inclination $i$.
If all residuals of the fit are smaller than $0.3$\arcsec, we regard the object as a candidate.

\subsection{Collection of similar candidates} 
The linking procedure can eliminate most of the false combinations. 
However, it is still possible that one track in a night can still link to two different tracks in the other night and produce
two valid candidates. 
It is difficult to distinguish the correct link from the false links without any visual inspection. 
Therefore, we collect the similar candidates (the candidates that share part of their tracks). 
These candidates are then visually inspected.

\subsection{Visual Inspection}
\label{sec:zoo}
\begin{figure}
\includegraphics[width = 1.\textwidth]{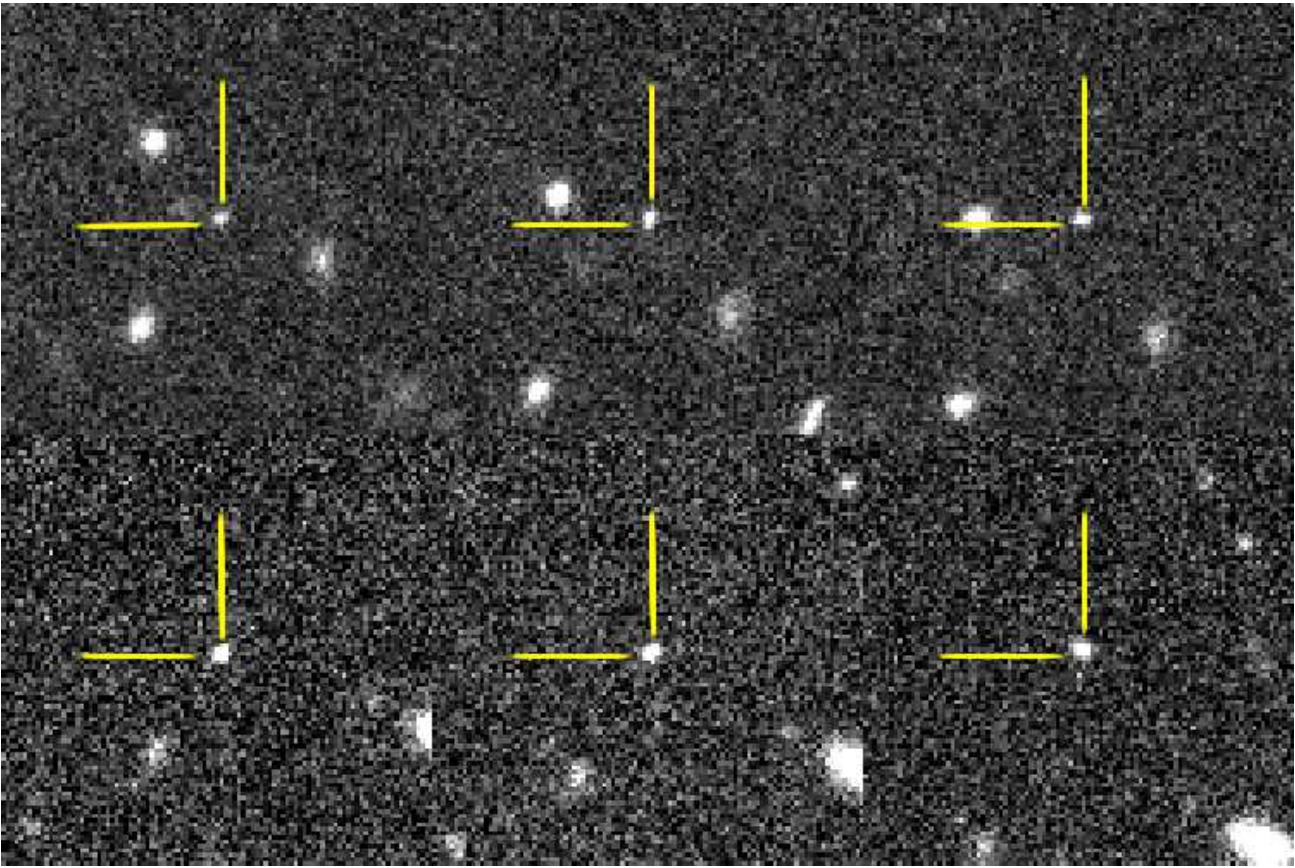}
\caption{Postage stamp images of a candidate object(purple lines indicate location of the object), assembled for visual
inspection on the Zooniverse platform. }\label{fig:cand}
\end{figure}

Visual inspection is frequently used to check whether moving object candidates are real objects or false positives.
The pipeline generates postage stamp images (fits and png) for each candidate, which are centered in the postage stamp images
(see Figure~\ref{fig:cand}).
We only inspect objects with $a > 20$~au (from the IOD orbit fit) in order to avoid including asteroids. 
A set of six truncated images (three images are from day one, three images are from day two) is uploaded to the {\tt
Zooniverse} platform for the inspection by team members.
{\tt Zooniverse} is a citizen science interface, which is used by many scientific projects, including astronomy and ecology
\citep{tro17}. The {\tt Zooniverse} project builder provides a very simple interface to launch a visual inspection project
with no required understanding the backbone structure. 
Because of the HSC-SSP data policy, we built the visual inspection page as a private project. 
Every candidate which is validated by at least two team members is listed as a real object.


\subsection{Process Note}
This pipeline simply searches the candidates moving within a single {\tt HEALPix} region. 
The current goal of this pipeline is to discover moving objects that move at a rate slow enough to keep most of them in the
same {\tt HEALPix} region within one observational run. 
The detection of objects that cross from one {\tt HEALPix} region to another will be future work for finding asteroids or
other nearby objects.

\section{Preliminary Results and Discussion} \label{sec:result}
We applied this pipeline to the HSC-SSP data taken from March 2014 to November 2015. 
Around two thousand moving-objects candidates in each dark run were detected, including asteroids, centaurs, and TNOs. 
In this paper, we only present the results of our TNO search, so only the objects with semi-major axes larger than 20~au were
selected for visual inspection. 
More than 500~candidates (including false positives) meet this criterion.
A total of 231~of these TNO candidates pass visual inspection by at least two members (see Table~\ref{tab:cands}).
Although the HSC-SSP had nine dark runs in first data release, we only analyzed the dark runs with solar elongations in our
required range, and which also have a sufficient number of exposures for the detection pipeline to function. 
These are images taken in 2014/11, 2015/03, 2015/05 and 2015/10 runs (See Table~\ref{tab:first}). 

\subsection{Detection Efficiency}
\begin{figure}
\includegraphics[width = 1.0\textwidth]{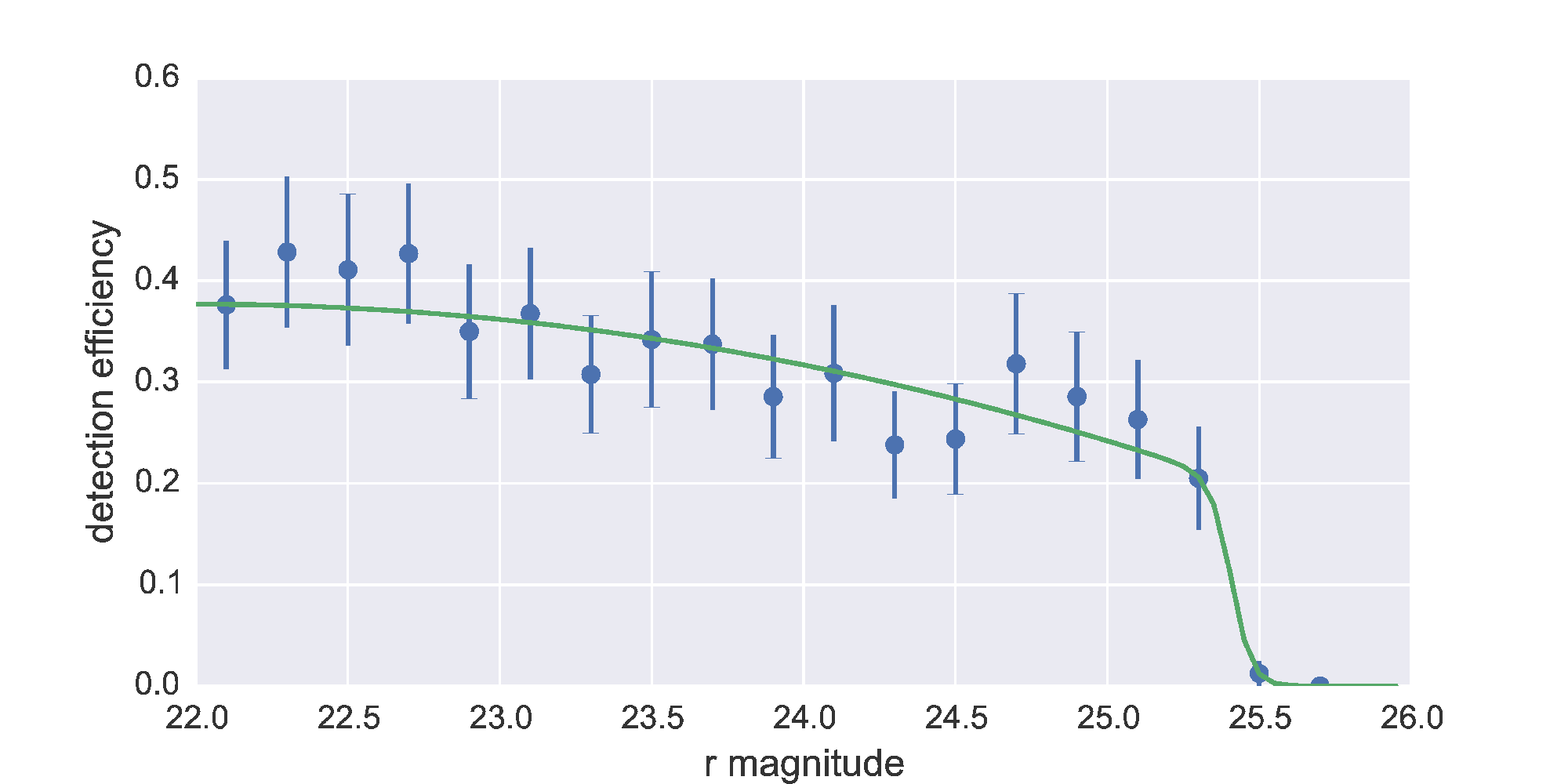} 
\caption{The detection efficiency for the testing {\tt HEALPix}, split into bins of magnitude. The best-fit detection
efficiency function (see Eq.~\ref{eq1}) is illustrated. The original synthetic moving objects have an uniform magnitude
distribution with range of 22 - 26.
}\label{fig:magf}
\end{figure}

To understand the survey depth and the detection efficiency (both as a function of object brightness and moving rate) of the
HSC-SSP, we generated synthetic moving objects with a magnitude range of 22 - 26 and a rate of motion of
1\arcsec--15\arcsec/hr (emulating objects at 10-100~au at opposition). 
Each chip has around 100~synthetic objects added. 
The PSFs of the synthetic moving objects are generated using a 2D~Gaussian function with the FWHM equal to the mean seeing. 
The raw images were implanted with the synthetic moving objects, and were then processed by {\tt hscPipe} in the same way as
the regular data processing to inspect the net efficiency of the detection pipeline.
The result is shown in Figure ~\ref{fig:magf}. The efficiency as a function of magnitude was fit with a detection efficiency
function in \citet{ale16} and \citet{ban16} as follows:

\begin{equation} \label{eq1}
\eta(m_r) = \eta_0-c(m_r-22)^2\left[1+\exp\left(\frac{m_r-m_L}{w}\right)\right]^{-1}
\end{equation}

This function gives an efficiency of about $\eta_0$ at $m_r = 22$, and it decreases quadratically with magnitude until it
drops more steeply near the magnitude limit $m_L$ with an exponential falloff width $w$. 
The efficiency of HSC-SSP is best represented with $\eta_0$ = 0.38, $c$ = 0.015, $m_L$ = 25.41 and $w$ = 0.033. 
Considering that the TNO detection criteria of our pipeline require at least three individual detections on two different
nights (six in total), the maximum detection efficiency below $40\%$ is expected.
Given at least six exposures and the $\sim85$\% to $90$\% fill factor in a single HSC exposure, the net efficiency will be $0.85^6 \sim 0.38$ to $0.9^6
\sim 0.53$ for detecting the moving objects.



\subsection{Orbit}
As discussed above, the orbital parameters and uncertainties of each candidate are calculated using the \citet{bk00} routine.
The orbital uncertainties of these candidates are not negligible because the observations generally span less than a month. 
Due to the short arc of these observations, the predicted on-sky position uncertainty grows quickly, making the objects
difficult to follow-up after the semester of discovery.
We then tested \citet{bk00} routine using known TNOs with a cadence similar to HSC-SSP, the results show that the orbital inclination is the only reliable orbital parameter for a TNO with short-arc observations. 
We are thus able to separate the sample \citep{volk11} into a cold population ($i + \sigma_i < 5\arcdeg$) and a hot population
($i - \sigma_i > 5\arcdeg$). 
After this process we find a total of 164 hot objects in our list of TNOs.
This simple cut means there will be a minor contamination of hot objects in the cold population, but our hot population is
useful for qualitative analysis.

\subsection{Absolute Magnitude Distribution}
\begin{figure}
\includegraphics[width = 1.0\textwidth]{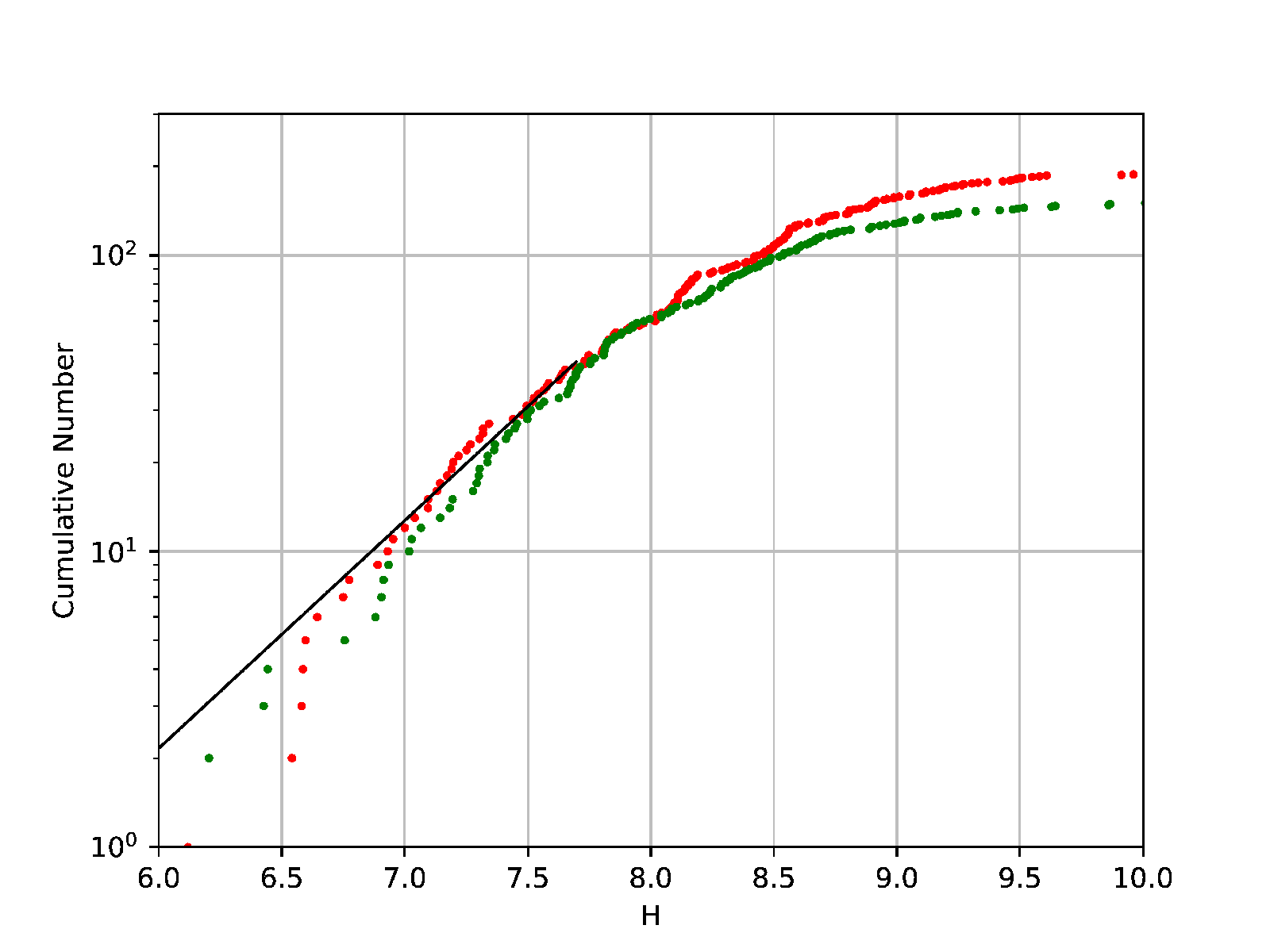} 
\caption{Absolute magnitude distributions of $r$-band (red points) and $g$-band (green points, shifted by a general color term
of 0.6) of the hot population KBOs. The two distributions have very similar shapes throughout the entire magnitude range. The
black line indicates the best-fit power-law distribution to the total magnitude distribution through $H = 7.7$, which has a
power-law slope of .  }\label{fig:lm}
\end{figure}

Although the orbital uncertainties of these candidates don't allow dynamical classification, flux corrections are still
achievable. The absolute magnitude $H$ of each candidate can be calculated using the photometry and initial orbital solutions.
Even though the short arcs of these candidates produce uncertainties of around 2~au in the barycentric distance by \citet{bk00}, our results are comparable with \citet{fra14}.
These preliminary results have not yet been debiased with the detection efficiencies, so we only consider the bright
candidates with $H_{r}~<~7.7$.
In Figure~\ref{fig:lm}, the best fit slope of a single power law to the absolute magnitude distribution of hot objects with $H_{r} < 7.7$ is
0.77. This value is consistent with the results from \citet{fra14}, which has a slope of 0.87$^{+0.07}_{-0.2}$ before the
break magnitude at $H_{r} = 7.7$.

\subsection{Color Distribution}
The bimodal color distribution of outer Solar System objects has been known for a decade
\citep{pei03, bar05, per10, pei12, lac14}.
This bimodal color distribution implies that different evolution processes or surface compositions may occur in the early
history of the outer Solar System.
While the selected targets in the literature include the observational biases, resulting in non-uniform samples,
recent research using the Subaru HSC resulted in a robust bimodal color distribution of small objects in the hot
population (\citet{wong17}, Figure 3).
The $g-i$ color distribution of our sample is compatible with these results.   
Although the rotation effects and observational biases have not removed from our results, the photometry from using two
filters in different nights could still provide an approximate color properties of the TNO surfaces.

\begin{figure}
\includegraphics[width = 1.0\textwidth]{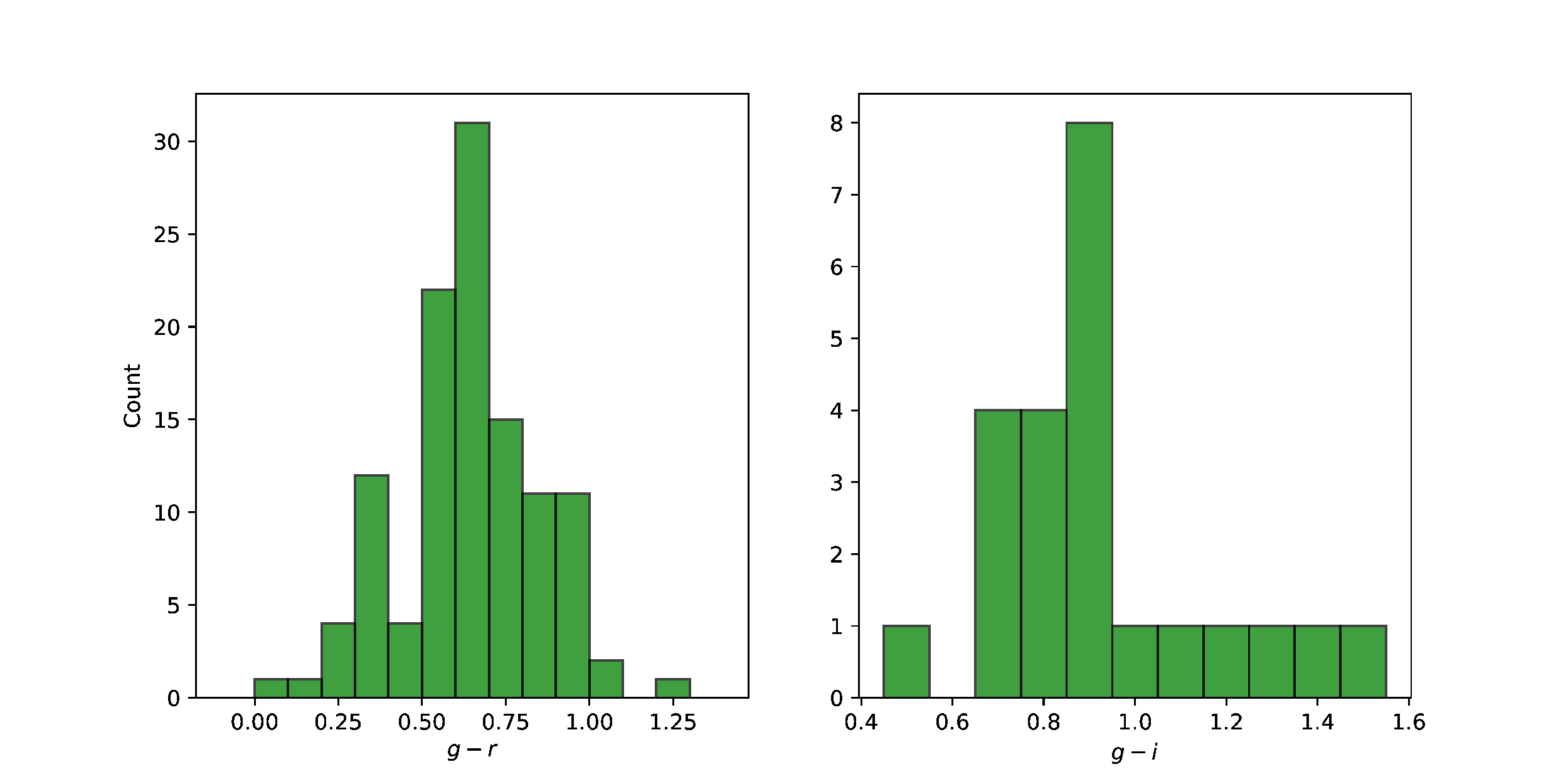}
\caption{The $g-r$ and $g-i$ color distribution of HSC-SSP TNOs. The highest bin of $g-i$ consists with the bluer peak of the bimodal color in
\citet{wong17}  }\label{fig:color}
\end{figure}

As mentioned in above, the HSC-SSP is not optimized for Solar System science, and this is reflected in the choice of filters
for different observations as well as the survey cadence.
Measurements of $g-r$ colors of 116 hot TNOs indicate a peak of count number at 0.6 (See Figure \ref{fig:color}).
Measurements of $g-i$ colors of only 24 hot TNOs are available in the first data release (images taken between March and May
2015). In Figure \ref{fig:color}, a sharp bluer peak at $g-i = 0.9$ is consistent
with the bluer peak ($g-i = 0.91$) of the bimodal color distribution in \citet{wong17}.
Unfortunately, the sample of redder TNOs ($g - i > 1.0$) we found with the HSC-SSP is currently too small to compare with the
redder peak at $g-i = 1.42$ reported by \citet{wong17}.
With additional discoveries of TNOs beyond the first HSC-SSP data release, a larger database of the hot TNO $g-i$
colors will be available to examine the bimodal color distribution and further explore the surface properties of these objects.


\subsection{Existence of vertical Centaur/TNO Belt?}
The recent discovery of a possible common plane in giant planet region provides 
evidence of the existence of a high-inclination ($60\arcdeg < i < 120\arcdeg$) and high-perihelion (q $>$ 10 AU) belt \citep{chen16}.
The clustering of ascending nodes is the significant feature of this common plane.
To farther confirm the existence of the vertical plane, the total numbers of 
high-inclination Centaurs/TNOs need to be increased for improving the significant in statistics.
The surface density of high-inclination Centaurs/TNOs ``on'' and ``off'' this common plane is a probe of intrinsic population. 
Although the survey fields of entire HSC-SSP almost have no overlapping region with proposed sky region of this belt (See Figure
\ref{fig:area} in this study and Figure 4 in \citet{chen16}), the HSC-SSP could provide the ``off'' plane observations as a
control samples to examine this hypothesis.

In our candidates, the highest inclination of prograde objects is $53\arcdeg$; none of the eight retrograde objects are in
this vertical plane ($60\arcdeg < i < 120\arcdeg$).
For a sanity check, we used another orbital fitting code, {\tt OpenOrb} \citep{gra09}, to validate the inclinations of the
retrograde objects.
The results show that two of the eight retrograde objects are possibly main-belt asteroids with low inclinations, and the other
six have the orbital solutions with the inclination near $\sim$30\arcdeg or $\sim$130\arcdeg.
Although both orbit fitting codes show the very low possibility that these eight retrograde objects are orbiting in this vertical plane,
additional follow-up observations definitely will provide a robust inclination solution.

The result of a non-detection of high-inclination TNOs in this vertical plane corresponds a low column density of ``off''
plane objects. However, for getting a relevant upper limit, the combination of the HSC-SSP results, follow-up observations and
future on-common plane observations using also the HSC will provide a statistically valid test of the existence of this
vertical Centaur/TNO Belt.



\subsection{Asteroids}
This pipeline has the ability to find the asteroid tracks in a single night, and
thousands of asteroids were detected in this data set.
However, the orbital fitting code we used was designed for identifying TNOs, and the orbit fitting for asteroids is not
robust. We will use an alternative code for asteroids in a future analysis.

\section{Summary} \label{sec:con}
The HSC-SSP is currently the deepest ongoing survey, and will cover more than 1400~deg$^2$. 
This data set provides an excellent opportunity to explore various topics in observational astronomy. 
In this study, we have described in detail the main algorithms of our moving object pipeline, which could be used in a survey
with dithering pointings, e.g. HSC-SSP.
This search algorithm could provide a better approach to discover Solar System objects in the future large surveys, which have
non-optimal cadences or dithered pointings. 
Compared with other moving object pipelines for searching for TNOs, this pipeline has fewer limitations on survey cadence and
pointing, which are very critical parameters for the detection of moving objects. 
The false positive rate of single detections in the catalog was significantly reduced to $15$\% with our ML algorithms. 
The ML also indirectly decreases the false positive rate of moving object candidates. 
We used the pipeline we have developed to examine the first data release of the HSC-SSP (2014-2015).

The preliminary search of this first data set has yielded 231 TNOs candidates that pass our search criteria. 
The lack of follow-up observations leads to large uncertainties in orbital parameters except for inclination.
The absolute magnitude distribution of HSC-SSP TNOs shows a slope of 0.77 which is shallower than a recently study of 1.45
\citep{wong17}, but our value is consistent with the study with proper calibrations (\citet{fra14}, $\alpha =
0.87^{+0.07}_{-0.2}$). 
The $g-i$ color distribution of HSC-SSP hot TNOs agrees with the bluer peak of bimodal distribution mentioned in the
recent study. 
More detections in future HSC-SSP data sets will provide a significant number of detections of faint TNOs which can be used to
improve our understanding of the formation history of the Solar System.

\begin{ack}
We are grateful to Yi-Ching Chiu, Paul Price and Hisanori Furusawa for kindly suggestions and helps in data process.
This work was supported in part by MOST Grant: MOST 104-2119-008-024 (TANGO II) and MOE under the Aim for Top University
Program NCU, and Macau Technical Fund: 017/2014/A1 and 039/2013/A2. HWL acknowledges the support of the CAS Fellowship for
Taiwan-Youth-Visiting-Scholars under the grant no. 2015TW2JB0001.
This publication uses data generated via the Zooniverse.org platform, development of is which funded by generous support,
including a Global Impact Award from Google, and by a grant from the Alfred P. Sloan Foundation.
The Hyper Suprime-Cam (HSC) collaboration includes the astronomical communities of Japan and Taiwan, and Princeton University.
The HSC instrumentation and software were developed by the National Astronomical Observatory of Japan (NAOJ), the Kavli 
Institute for the Physics and Mathematics of the Universe (Kavli IPMU), the University of Tokyo, the High Energy Accelerator
Research Organization (KEK), the Academia Sinica Institute for Astronomy and Astrophysics in Taiwan (ASIAA), and Princeton
University. Funding was contributed by the FIRST program from Japanese Cabinet Office, the Ministry of Education, Culture, 
Sports, Science and Technology (MEXT), the Japan Society for the Promotion of Science (JSPS), Japan Science and Technology
Agency (JST), the Toray Science Foundation, NAOJ, Kavli IPMU, KEK, ASIAA, and Princeton University. 

This paper makes use of software developed for the Large Synoptic Survey Telescope. We thank the LSST Project for making their
code available as free software at  http://dm.lsst.org

The Pan-STARRS1 Surveys (PS1) have been made possible through contributions of the Institute for Astronomy, the University of
Hawaii, the Pan-STARRS Project Office, the Max-Planck Society and its participating institutes, the Max Planck Institute for
Astronomy, Heidelberg and the Max Planck Institute for Extraterrestrial Physics, Garching, The Johns Hopkins University,
Durham University, the University of Edinburgh, Queen’s University Belfast, the Harvard-Smithsonian Center for Astrophysics,
the Las Cumbres Observatory Global Telescope Network Incorporated, the National Central University of Taiwan, the Space 
Telescope Science Institute, the National Aeronautics and Space Administration under Grant No. NNX08AR22G issued through the
Planetary Science Division of the NASA Science Mission Directorate, the National Science Foundation under Grant No. 
AST-1238877, the University of Maryland, and Eotvos Lorand University (ELTE) and the Los Alamos National Laboratory.

Based on data collected at the Subaru Telescope and retrieved from the HSC data archive system, which is operated by Subaru
Telescope and Astronomy Data Center, National Astronomical Observatory of Japan.
\end{ack}


\newpage


\newpage



\begin{table}
  \tbl{The target regions in the first data release of HSC-SSP used in the moving object search.}{%
  \begin{tabular*}{\textwidth}{c @{\extracolsep{\fill}}ccccccc}
      \hline
      Observation Date &  R.A. & Decl. & Ec. lat. & Effective Area & Filter & Number of Candidates\\
                       & (deg) & (deg) & (deg)    & deg$^2$        &      &  \\
      \hline
      2014/11 & 34.5  & -5.4 & -18.1 & 38.5 & $g, r, i$ & 8\\
      2015/03 & 180.2 & 0.1  &  0.2  & 29.7 & $g, r, i$ & 92\\
      2015/03 & 217.3 & 0.2  & 14.1  & 24.6 & $g, r, i$ & 9\\
      2015/05 & 218.5 & -1.2 &  13.2 & 29.3 & $g, r, i$ & 28\\
      2015/05 & 245.5 & 43.1 &  63.0 & 11.1 & $g, r, i$ & 0\\
      2015/10 & 27.7  & -1.1 & -11.7 & 79.2 & $g, r$ & 93\\
      2015/10 & 340.4 & 0.5  &  8.1  & 8.6 & $g, r$ & 1\\
      \hline
    \end{tabular*}}\label{tab:first}
\end{table}


\begin{longtable}{ccccccc}
\caption{Orbital parameters and photometry of HSC-SSP TNOs in 2014 and 2015.} \label{tab:cands} \\
\hline              
  Name & $a$ (au) & $e$ & $i$ (deg) & g (mag) & r (mag) & i (mag)  \\

\endfirsthead
\endhead
  \hline
\endfoot
  \hline
\endlastfoot
  \hline

20141113-1127.cand001  &  41.2$\pm$20.9  &  0.02$\pm$0.58  &  27.45$\pm$3.2  &                           &  24.52$\pm$0.08  &24.49$\pm$0.21\\
20141113-1127.cand002  &  40.3$\pm$20.9  &  0.02$\pm$0.58  &  16.70$\pm$0.5  &                           &  23.73$\pm$0.02  &23.55$\pm$0.09\\
20141113-1127.cand003  &  48.3$\pm$24.7  &  0.02$\pm$0.57  &  19.48$\pm$0.5  &                           &  24.27$\pm$0.06  &23.81$\pm$0.17\\
20141113-1127.cand004  &  40.0$\pm$20.2  &  0.01$\pm$0.58  &  25.92$\pm$2.8  &                           &  24.50$\pm$0.12  &24.04$\pm$0.10\\
20141113-1127.cand005  &  45.8$\pm$23.5  &  0.02$\pm$0.58  &  38.23$\pm$9.1  &  22.23$\pm$0.02  &                           &21.87$\pm$0.03\\
20141113-1127.cand006  &  40.8$\pm$21.1  &  0.03$\pm$0.58  &  19.93$\pm$0.3  &  24.87$\pm$0.04  &                           &23.90$\pm$0.09\\
20141113-1127.cand007  &  36.5$\pm$19.0  &  0.04$\pm$0.57  &  19.75$\pm$0.5  &                           &  23.97$\pm$0.00  &23.79$\pm$0.08\\
20141113-1127.cand008  &  37.1$\pm$19.4  &  0.03$\pm$0.58  &  21.52$\pm$0.2  &                           &  24.03$\pm$0.04  &23.79$\pm$0.16\\
20150312-0330.cand001  &  45.3$\pm$23.5  &  0.02$\pm$0.56  &  16.78$\pm$6.2  &                           &  24.23$\pm$0.03  &23.96$\pm$0.10\\
20150312-0330.cand002  &  44.7$\pm$23.2  &  0.02$\pm$0.56  &  2.69$\pm$1.0  &  25.69$\pm$0.07  &                           &24.23$\pm$0.02\\
20150312-0330.cand003  &  42.9$\pm$22.3  &  0.02$\pm$0.56  &  1.01$\pm$0.4  &  25.46$\pm$0.10  &                           &24.19$\pm$0.10\\
20150312-0330.cand004  &  44.9$\pm$23.3  &  0.02$\pm$0.56  &  0.33$\pm$0.4  &                           &  24.53$\pm$0.10  &24.06$\pm$0.11\\
20150312-0330.cand005  &  41.9$\pm$21.8  &  0.02$\pm$0.56  &  1.69$\pm$0.7  &                           &  25.17$\pm$0.19  &24.83$\pm$0.22\\
20150312-0330.cand006  &  44.5$\pm$23.1  &  0.02$\pm$0.56  &  10.80$\pm$3.9  &                           &  23.92$\pm$0.02  &23.94$\pm$0.32\\
20150312-0330.cand007  &  44.9$\pm$23.3  &  0.02$\pm$0.56  &  0.40$\pm$0.2  &                           &  25.05$\pm$0.08  &24.55$\pm$0.11\\
20150312-0330.cand008  &  46.3$\pm$24.0  &  0.02$\pm$0.56  &  0.47$\pm$0.5  &                           &  25.71$\pm$0.02  &25.24$\pm$0.09\\
20150312-0330.cand009  &  45.9$\pm$23.8  &  0.02$\pm$0.56  &  2.92$\pm$1.1  &                           &  25.07$\pm$0.11  &24.79$\pm$0.17\\
20150312-0330.cand010  &  44.1$\pm$22.9  &  0.02$\pm$0.56  &  1.44$\pm$0.7  &                           &  24.82$\pm$0.10  &24.30$\pm$0.06\\
20150312-0330.cand011  &  44.5$\pm$23.2  &  0.02$\pm$0.58  &  31.43$\pm$13.4  &  25.43$\pm$0.11  &  24.87$\pm$0.05  &                       \\
20150312-0330.cand012  &  41.4$\pm$21.7  &  0.02$\pm$0.58  &  29.38$\pm$12.5  &  24.26$\pm$0.06  &                           &23.35$\pm$0.06\\
20150312-0330.cand013  &  39.6$\pm$20.7  &  0.02$\pm$0.57  &  3.99$\pm$1.2  &  25.50$\pm$0.09  &                           &24.07$\pm$0.05\\
20150312-0330.cand014  &  47.6$\pm$24.7  &  0.02$\pm$0.57  &  25.84$\pm$10.2  &  24.16$\pm$0.07  &                           &23.43$\pm$0.05\\
20150312-0330.cand015  &  44.7$\pm$23.2  &  0.02$\pm$0.56  &  2.30$\pm$0.2  &                           &  24.46$\pm$0.06  &23.98$\pm$0.15\\
20150312-0330.cand016  &  41.3$\pm$21.6  &  0.02$\pm$0.57  &  20.52$\pm$8.0  &                           &  25.46$\pm$0.15  &24.74$\pm$0.25\\
20150312-0330.cand017  &  46.6$\pm$24.1  &  0.02$\pm$0.56  &  2.75$\pm$0.8  &                           &  25.13$\pm$0.13  &24.55$\pm$0.09\\
20150312-0330.cand018  &  44.2$\pm$23.0  &  0.02$\pm$0.56  &  1.98$\pm$0.4  &                           &  25.17$\pm$0.21  &24.65$\pm$0.04\\
20150312-0330.cand019  &  41.5$\pm$22.1  &  0.02$\pm$0.63  &  48.99$\pm$31.6  &                           &  24.43$\pm$0.05  &24.06$\pm$0.05\\
20150312-0330.cand020  &  41.4$\pm$21.7  &  0.02$\pm$0.58  &  29.37$\pm$12.4  &                           &  23.59$\pm$0.04  &23.44$\pm$0.03\\
20150312-0330.cand021  &  46.1$\pm$23.9  &  0.02$\pm$0.56  &  1.81$\pm$0.6  &  24.50$\pm$0.06  &                           &23.45$\pm$0.03\\
20150312-0330.cand022  &  41.2$\pm$21.5  &  0.02$\pm$0.57  &  17.88$\pm$6.8  &  24.83$\pm$0.07  &                           &23.89$\pm$0.11\\
20150312-0330.cand023  &  54.7$\pm$28.1  &  0.02$\pm$0.55  &  6.42$\pm$2.1  &  25.16$\pm$0.14  &                           &23.78$\pm$0.04\\
20150312-0330.cand024  &  41.9$\pm$21.9  &  0.02$\pm$0.56  &  1.17$\pm$0.3  &  24.89$\pm$0.12  &                           &23.63$\pm$0.10\\
20150312-0330.cand025  &  43.9$\pm$23.0  &  0.02$\pm$0.59  &  38.57$\pm$18.5  &  25.41$\pm$0.24  &                           &24.62$\pm$0.25\\
20150312-0330.cand026  &  44.5$\pm$23.1  &  0.02$\pm$0.56  &  1.56$\pm$0.2  &  24.85$\pm$0.18  &                           &23.38$\pm$0.06\\
20150312-0330.cand027  &  48.6$\pm$25.1  &  0.02$\pm$0.56  &  1.50$\pm$0.1  &  23.77$\pm$0.05  &                           &22.52$\pm$0.15\\
20150312-0330.cand028  &  42.0$\pm$21.9  &  0.02$\pm$0.56  &  1.27$\pm$0.2  &  24.23$\pm$0.23  &  23.02$\pm$0.02  &                       \\
20150312-0330.cand029  &  44.5$\pm$23.1  &  0.02$\pm$0.56  &  7.61$\pm$2.7  &                           &                           &24.11$\pm$0.09\\
20150312-0330.cand030  &  50.2$\pm$25.9  &  0.02$\pm$0.55  &  1.76$\pm$0.6  &                           &  24.93$\pm$0.11  &24.86$\pm$0.12\\
20150312-0330.cand031  &  43.8$\pm$22.8  &  0.02$\pm$0.56  &  2.14$\pm$0.4  &                           &  23.41$\pm$0.07  &22.90$\pm$0.01\\
20150312-0330.cand032  &  54.2$\pm$27.8  &  0.02$\pm$0.55  &  1.70$\pm$0.8  &                           &  25.36$\pm$0.16  &24.69$\pm$0.07\\
20150312-0330.cand033  &  42.2$\pm$22.0  &  0.02$\pm$0.56  &  2.91$\pm$1.1  &                           &  24.12$\pm$0.10  &23.74$\pm$0.10\\
20150312-0330.cand034  &  42.3$\pm$22.0  &  0.02$\pm$0.56  &  4.76$\pm$1.7  &                           &  24.69$\pm$0.12  &24.22$\pm$0.07\\
20150312-0330.cand035  &  46.6$\pm$24.1  &  0.02$\pm$0.56  &  4.96$\pm$1.7  &                           &  23.15$\pm$0.11  &22.68$\pm$0.05\\
20150312-0330.cand036  &  43.5$\pm$22.6  &  0.02$\pm$0.56  &  2.60$\pm$0.9  &                           &  25.14$\pm$0.14  &24.64$\pm$0.09\\
20150312-0330.cand037  &  43.1$\pm$22.5  &  0.02$\pm$0.56  &  3.62$\pm$1.3  &                           &  24.00$\pm$0.03  &23.57$\pm$0.04\\
20150312-0330.cand038  &  47.4$\pm$24.5  &  0.02$\pm$0.56  &  1.00$\pm$0.0  &                           &  25.32$\pm$0.22  &24.81$\pm$0.18\\
20150312-0330.cand039  &  47.6$\pm$24.6  &  0.02$\pm$0.56  &  4.01$\pm$1.4  &                           &  24.65$\pm$0.07  &24.32$\pm$0.05\\
20150312-0330.cand040  &  43.5$\pm$22.6  &  0.02$\pm$0.56  &  1.32$\pm$0.5  &                           &  25.12$\pm$0.12  &24.51$\pm$0.11\\
20150312-0330.cand041  &  42.2$\pm$22.0  &  0.02$\pm$0.56  &  1.49$\pm$0.3  &                           &  25.08$\pm$0.08  &24.79$\pm$0.16\\
20150312-0330.cand042  &  36.0$\pm$19.0  &  0.03$\pm$0.58  &  24.77$\pm$10.4  &                           &  25.29$\pm$0.03  &24.90$\pm$0.03\\
20150312-0330.cand043  &  47.8$\pm$24.8  &  0.02$\pm$0.57  &  32.71$\pm$13.9  &                           &  25.16$\pm$0.08  &24.83$\pm$0.04\\
20150312-0330.cand044  &  44.5$\pm$23.1  &  0.02$\pm$0.56  &  0.85$\pm$0.2  &                           &  25.28$\pm$0.07  &24.90$\pm$0.12\\
20150312-0330.cand045  &  47.8$\pm$24.7  &  0.02$\pm$0.56  &  1.18$\pm$0.1  &                           &  24.83$\pm$0.07  &24.43$\pm$0.04\\
20150312-0330.cand046  &  46.1$\pm$23.9  &  0.02$\pm$0.56  &  1.70$\pm$0.6  &                           &  24.64$\pm$0.12  &24.37$\pm$0.07\\
20150312-0330.cand047  &  48.2$\pm$24.9  &  0.02$\pm$0.56  &  1.05$\pm$0.4  &                           &  23.45$\pm$0.03  &22.96$\pm$0.03\\
20150312-0330.cand048  &  53.1$\pm$27.3  &  0.02$\pm$0.55  &  8.49$\pm$3.0  &                           &  24.96$\pm$0.04  &24.81$\pm$0.12\\
20150312-0330.cand049  &  43.6$\pm$22.7  &  0.02$\pm$0.56  &  1.20$\pm$0.3  &                           &  24.78$\pm$0.03  &24.34$\pm$0.07\\
20150312-0330.cand050  &  42.5$\pm$22.2  &  0.02$\pm$0.56  &  1.26$\pm$0.4  &                           &  25.04$\pm$0.03  &24.46$\pm$0.16\\
20150312-0330.cand051  &  45.5$\pm$23.6  &  0.02$\pm$0.56  &  1.22$\pm$0.2  &                           &  25.14$\pm$0.04  &24.93$\pm$0.06\\
20150312-0330.cand052  &  38.3$\pm$20.1  &  0.03$\pm$0.57  &  6.72$\pm$2.5  &                           &  25.17$\pm$0.15  &24.62$\pm$0.17\\
20150312-0330.cand053  &  42.4$\pm$22.1  &  0.02$\pm$0.56  &  1.24$\pm$0.2  &                           &  25.02$\pm$0.16  &24.51$\pm$0.08\\
20150312-0330.cand054  &  43.4$\pm$22.6  &  0.02$\pm$0.56  &  2.25$\pm$0.8  &                           &  25.37$\pm$0.06  &24.54$\pm$0.06\\
20150312-0330.cand055  &  46.7$\pm$24.2  &  0.02$\pm$0.56  &  3.22$\pm$1.2  &                           &  23.72$\pm$0.05  &23.56$\pm$0.04\\
20150312-0330.cand056  &  42.3$\pm$22.2  &  0.02$\pm$0.59  &  38.91$\pm$19.1  &                           &  24.67$\pm$0.08  &24.31$\pm$0.08\\
20150312-0330.cand057  &  56.0$\pm$28.8  &  0.02$\pm$0.55  &  20.77$\pm$7.5  &  25.73$\pm$0.14  &                           &24.44$\pm$0.07\\
20150312-0330.cand058  &  40.2$\pm$21.0  &  0.02$\pm$0.58  &  24.49$\pm$9.9  &  24.47$\pm$0.08  &                           &23.62$\pm$0.07\\
20150312-0330.cand059  &  37.4$\pm$19.8  &  0.03$\pm$0.60  &  35.30$\pm$17.0  &  24.74$\pm$0.13  &                           &23.83$\pm$0.07\\
20150312-0330.cand060  &  45.4$\pm$23.6  &  0.02$\pm$0.56  &  22.11$\pm$8.5  &  25.58$\pm$0.05  &                           &25.12$\pm$0.37\\
20150312-0330.cand061  &  40.4$\pm$21.1  &  0.02$\pm$0.57  &  1.68$\pm$0.1  &  25.10$\pm$0.08  &                           &24.50$\pm$0.12\\
20150312-0330.cand062  &  43.4$\pm$22.6  &  0.02$\pm$0.56  &  1.45$\pm$0.3  &  25.19$\pm$0.19  &  24.37$\pm$0.11  &                       \\
20150312-0330.cand063  &  38.8$\pm$20.4  &  0.03$\pm$0.57  &  14.87$\pm$5.7  &  25.64$\pm$0.25  &  25.05$\pm$0.08  &                       \\
20150312-0330.cand064  &  42.9$\pm$22.3  &  0.02$\pm$0.56  &  2.19$\pm$0.8  &                           &  24.87$\pm$0.07  &24.48$\pm$0.08\\
20150312-0330.cand065  &  44.1$\pm$22.9  &  0.02$\pm$0.56  &  5.26$\pm$1.8  &                           &  25.10$\pm$0.17  &24.56$\pm$0.07\\
20150312-0330.cand066  &  37.4$\pm$19.8  &  0.03$\pm$0.60  &  35.20$\pm$16.9  &                           &  24.04$\pm$0.04  &23.86$\pm$0.03\\
20150312-0330.cand067  &  41.8$\pm$21.8  &  0.02$\pm$0.56  &  7.25$\pm$2.5  &                           &  24.95$\pm$0.17  &24.81$\pm$0.03\\
20150312-0330.cand068  &  46.2$\pm$23.9  &  0.02$\pm$0.56  &  2.82$\pm$1.0  &                           &  24.14$\pm$0.05  &23.84$\pm$0.03\\
20150312-0330.cand069  &  45.8$\pm$23.7  &  0.02$\pm$0.56  &  9.26$\pm$3.3  &                           &  25.03$\pm$0.11  &24.51$\pm$0.03\\
20150312-0330.cand070  &  44.2$\pm$23.0  &  0.02$\pm$0.56  &  1.10$\pm$0.3  &                           &  25.25$\pm$0.23  &24.98$\pm$0.22\\
20150312-0330.cand071  &  36.9$\pm$19.4  &  0.03$\pm$0.57  &  8.10$\pm$3.1  &                           &  24.13$\pm$0.09  &23.99$\pm$0.03\\
20150312-0330.cand072  &  43.8$\pm$22.8  &  0.02$\pm$0.56  &  0.85$\pm$0.2  &                           &  25.32$\pm$0.02  &24.91$\pm$0.14\\
20150312-0330.cand073  &  44.8$\pm$23.2  &  0.02$\pm$0.56  &  1.03$\pm$0.5  &                           &  24.92$\pm$0.11  &24.52$\pm$0.10\\
20150312-0330.cand074  &  44.3$\pm$23.0  &  0.02$\pm$0.56  &  1.66$\pm$0.7  &                           &  25.31$\pm$0.12  &24.40$\pm$0.82\\
20150312-0330.cand075  &  50.0$\pm$25.8  &  0.02$\pm$0.56  &  24.61$\pm$9.5  &                           &  25.20$\pm$0.12  &25.13$\pm$0.16\\
20150312-0330.cand076  &  47.9$\pm$24.8  &  0.02$\pm$0.56  &  1.55$\pm$0.6  &                           &  24.29$\pm$0.06  &23.97$\pm$0.13\\
20150312-0330.cand077  &  49.7$\pm$25.6  &  0.02$\pm$0.55  &  4.53$\pm$1.7  &                           &  25.22$\pm$0.15  &25.03$\pm$0.08\\
20150312-0330.cand078  &  44.1$\pm$22.9  &  0.02$\pm$0.56  &  1.90$\pm$0.2  &                           &  24.77$\pm$0.08  &24.59$\pm$0.17\\
20150312-0330.cand079  &  27.8$\pm$2.1  &  0.65$\pm$0.03  &  52.05$\pm$21.2  &  25.27$\pm$0.08  &                           &24.41$\pm$0.16\\
20150312-0330.cand080  &  41.3$\pm$21.6  &  0.02$\pm$0.58  &  28.05$\pm$11.7  &                           &  25.06$\pm$0.06  &24.56$\pm$0.06\\
20150312-0330.cand081  &  40.0$\pm$20.9  &  0.03$\pm$0.57  &  4.66$\pm$1.5  &  25.10$\pm$0.07  &  24.65$\pm$0.06  &                       \\
20150312-0330.cand082  &  49.3$\pm$25.4  &  0.02$\pm$0.55  &  3.24$\pm$0.9  &  25.56$\pm$0.10  &                           &24.32$\pm$0.09\\
20150312-0330.cand083  &  41.9$\pm$21.9  &  0.02$\pm$0.56  &  2.26$\pm$0.4  &  25.27$\pm$0.09  &                           &24.01$\pm$0.03\\
20150312-0330.cand084  &  42.5$\pm$22.1  &  0.02$\pm$0.56  &  2.90$\pm$0.7  &                           &  24.73$\pm$0.10  &24.23$\pm$0.05\\
20150312-0330.cand085  &  45.7$\pm$23.7  &  0.02$\pm$0.56  &  2.56$\pm$0.6  &                           &  23.41$\pm$0.01  &23.08$\pm$0.04\\
20150312-0330.cand086  &  40.6$\pm$21.2  &  0.02$\pm$0.57  &  6.64$\pm$2.2  &                           &  25.42$\pm$0.04  &24.99$\pm$0.13\\
20150312-0330.cand087  &  40.5$\pm$21.2  &  0.02$\pm$0.57  &  11.23$\pm$4.1  &                           &  25.28$\pm$0.06  &24.78$\pm$0.16\\
20150312-0330.cand088  &  44.8$\pm$23.3  &  0.02$\pm$0.56  &  0.54$\pm$0.2  &  24.64$\pm$0.09  &  23.68$\pm$0.07  &                       \\
20150312-0330.cand089  &  44.1$\pm$22.9  &  0.02$\pm$0.56  &  1.86$\pm$0.8  &                           &  24.68$\pm$0.04  &24.31$\pm$0.03\\
20150312-0330.cand090  &  42.1$\pm$22.0  &  0.02$\pm$0.58  &  27.91$\pm$11.7  &                           &  25.26$\pm$0.12  &24.68$\pm$0.11\\
20150312-0330.cand091  &  45.6$\pm$23.6  &  0.02$\pm$0.56  &  0.90$\pm$0.5  &                           &  24.11$\pm$0.05  &23.65$\pm$0.07\\
20150312-0330.cand092  &  42.0$\pm$21.9  &  0.02$\pm$0.56  &  8.03$\pm$3.0  &                           &  25.50$\pm$0.01  &25.17$\pm$0.12\\
20150312-0330.cand093  &  42.5$\pm$22.1  &  0.02$\pm$0.59  &  25.94$\pm$12.2  &  25.74$\pm$0.08  &  25.27$\pm$0.19  &                       \\
20150312-0330.cand094  &  34.0$\pm$17.9  &  0.03$\pm$0.59  &  16.87$\pm$1.2  &                           &  23.69$\pm$0.07  &23.65$\pm$0.25\\
20150312-0330.cand095  &  36.6$\pm$19.1  &  0.03$\pm$0.59  &  22.73$\pm$3.2  &  25.32$\pm$0.10  &  24.67$\pm$0.11  &                       \\
20150312-0330.cand096  &  22.1$\pm$11.8  &  0.02$\pm$0.59  &  164.60$\pm$0.2  &  25.40$\pm$0.09  &  24.41$\pm$0.47  &                       \\
20150312-0330.cand097  &  43.3$\pm$22.3  &  0.02$\pm$0.58  &  15.93$\pm$0.1  &  24.56$\pm$0.06  &  24.21$\pm$0.03  &                       \\
20150312-0330.cand098  &  42.7$\pm$22.0  &  0.02$\pm$0.58  &  20.06$\pm$1.0  &  25.06$\pm$0.08  &  24.46$\pm$0.23  &                       \\
20150312-0330.cand099  &  28.4$\pm$0.4  &  0.93$\pm$0.70  &  12.32$\pm$3.6  &  24.36$\pm$0.10  &  24.92$\pm$0.04  &                       \\
20150312-0330.cand100  &  22.9$\pm$12.8  &  0.04$\pm$0.64  &  14.12$\pm$0.7  &  23.10$\pm$0.04  &                           &21.68$\pm$0.04\\
20150312-0330.cand101  &  42.4$\pm$21.9  &  0.02$\pm$0.58  &  25.51$\pm$3.7  &                           &                           &24.17$\pm$0.08\\
20150511-0525.cand001  &  24.1$\pm$13.3  &  0.04$\pm$0.48  &  121.98$\pm$10.9  &  25.12$\pm$0.06  &  24.36$\pm$0.08  &                       \\
20150511-0525.cand002  &  37.4$\pm$19.7  &  0.03$\pm$0.59  &  26.51$\pm$12.0  &  24.40$\pm$0.10  &  23.75$\pm$0.07  &                       \\
20150511-0525.cand003  &  45.7$\pm$23.6  &  0.02$\pm$0.57  &  12.25$\pm$0.9  &  25.47$\pm$0.06  &  24.79$\pm$0.09  &                       \\
20150511-0525.cand004  &  45.1$\pm$23.5  &  0.02$\pm$0.60  &  36.19$\pm$18.9  &  24.19$\pm$0.08  &  23.60$\pm$0.05  &                       \\
20150511-0525.cand005  &  35.4$\pm$18.8  &  0.03$\pm$0.60  &  25.12$\pm$10.7  &  24.80$\pm$0.39  &  24.02$\pm$0.11  &                       \\
20150511-0525.cand006  &  37.8$\pm$19.8  &  0.03$\pm$0.58  &  12.30$\pm$0.6  &  24.61$\pm$0.08  &  23.56$\pm$0.03  &                       \\
20150511-0525.cand007  &  44.6$\pm$23.1  &  0.02$\pm$0.57  &  15.97$\pm$3.3  &  25.33$\pm$0.16  &  24.45$\pm$0.06  &                       \\
20150511-0525.cand008  &  41.9$\pm$21.9  &  0.02$\pm$0.59  &  28.99$\pm$12.7  &  25.16$\pm$0.12  &  24.53$\pm$0.13  &                       \\
20150511-0525.cand009  &  26.9$\pm$14.7  &  0.04$\pm$0.62  &  25.19$\pm$7.1  &  24.66$\pm$0.08  &  24.01$\pm$0.07  &                       \\
20150511-0525.cand010  &  38.0$\pm$19.9  &  0.03$\pm$0.58  &  17.00$\pm$4.7  &  25.45$\pm$0.06  &  24.63$\pm$0.13  &                       \\
20150511-0525.cand011  &  39.5$\pm$20.7  &  0.03$\pm$0.58  &  22.50$\pm$8.0  &  25.30$\pm$0.01  &  24.69$\pm$0.14  &                       \\
20150511-0525.cand012  &  33.6$\pm$17.8  &  0.03$\pm$0.59  &  18.78$\pm$6.2  &  24.40$\pm$0.15  &  23.51$\pm$0.04  &                       \\
20150511-0525.cand013  &  41.1$\pm$21.4  &  0.02$\pm$0.57  &  13.81$\pm$0.5  &  25.22$\pm$0.15  &  24.41$\pm$0.06  &                       \\
20150511-0525.cand014  &  36.7$\pm$19.3  &  0.03$\pm$0.58  &  12.07$\pm$0.1  &  23.88$\pm$0.13  &  23.56$\pm$0.04  &                       \\
20150511-0525.cand015  &  39.6$\pm$20.7  &  0.03$\pm$0.57  &  14.82$\pm$0.5  &  24.10$\pm$0.04  &  23.56$\pm$0.05  &                       \\
20150511-0525.cand016  &  47.1$\pm$24.3  &  0.02$\pm$0.56  &  23.79$\pm$4.0  &  24.22$\pm$0.09  &                           &23.55$\pm$0.04\\
20150511-0525.cand017  &  40.2$\pm$20.9  &  0.03$\pm$0.57  &  15.09$\pm$0.3  &  24.30$\pm$0.05  &                           &23.14$\pm$0.02\\
20150511-0525.cand018  &  35.2$\pm$18.6  &  0.03$\pm$0.58  &  15.18$\pm$0.1  &  22.79$\pm$0.10  &                           &21.25$\pm$0.05\\
20150511-0525.cand019  &  41.8$\pm$21.8  &  0.02$\pm$0.57  &  21.06$\pm$5.7  &  24.00$\pm$0.08  &  23.25$\pm$0.05  &                       \\
20150511-0525.cand020  &  33.8$\pm$18.0  &  0.03$\pm$0.60  &  24.35$\pm$9.3  &  24.39$\pm$0.04  &                           &23.45$\pm$0.08\\
20150511-0525.cand021  &  35.8$\pm$18.9  &  0.03$\pm$0.58  &  17.94$\pm$2.1  &  25.04$\pm$0.13  &                           &24.31$\pm$0.16\\
20150511-0525.cand022  &  31.6$\pm$16.9  &  0.03$\pm$0.59  &  18.05$\pm$4.3  &  24.86$\pm$0.07  &                           &24.01$\pm$0.10\\
20150511-0525.cand023  &  37.8$\pm$19.8  &  0.03$\pm$0.57  &  14.95$\pm$0.1  &  24.63$\pm$0.09  &                           &23.49$\pm$0.04\\
20150511-0525.cand024  &  40.2$\pm$21.0  &  0.02$\pm$0.57  &  14.41$\pm$0.1  &  24.18$\pm$0.13  &                           &23.41$\pm$0.06\\
20150511-0525.cand025  &  25.3$\pm$13.7  &  0.03$\pm$0.50  &  160.92$\pm$1.1  &  24.28$\pm$0.05  &                           &23.58$\pm$0.07\\
20150511-0525.cand026  &  49.7$\pm$25.6  &  0.02$\pm$0.57  &  19.76$\pm$5.1  &  25.39$\pm$0.13  &                           &24.50$\pm$0.04\\
20150511-0525.cand027  &  40.4$\pm$21.1  &  0.02$\pm$0.59  &  28.69$\pm$11.5  &  23.70$\pm$0.08  &                           &22.80$\pm$0.03\\
20150511-0525.cand028  &  35.6$\pm$18.8  &  0.03$\pm$0.59  &  23.34$\pm$7.9  &  24.55$\pm$0.12  &                           &23.78$\pm$0.04\\
20151005-1020.cand001  &  46.9$\pm$24.3  &  0.02$\pm$0.56  &  6.92$\pm$0.4  &  25.01$\pm$0.09  &  24.30$\pm$0.07  &                       \\
20151005-1020.cand002  &  33.9$\pm$18.0  &  0.03$\pm$0.58  &  9.82$\pm$2.6  &  24.95$\pm$0.20  &  24.57$\pm$0.02  &                       \\
20151005-1020.cand003  &  38.9$\pm$20.4  &  0.02$\pm$0.57  &  6.71$\pm$0.4  &  24.41$\pm$0.02  &  23.76$\pm$0.03  &                       \\
20151005-1020.cand004  &  41.8$\pm$21.8  &  0.02$\pm$0.57  &  11.49$\pm$2.8  &  25.17$\pm$0.01  &  24.88$\pm$0.15  &                       \\
20151005-1020.cand005  &  41.9$\pm$21.8  &  0.02$\pm$0.57  &  12.65$\pm$3.8  &  25.15$\pm$0.16  &  24.17$\pm$0.07  &                       \\
20151005-1020.cand006  &  29.6$\pm$16.1  &  0.03$\pm$0.49  &  135.76$\pm$9.1  &  24.60$\pm$0.05  &  24.07$\pm$0.01  &                       \\
20151005-1020.cand007  &  43.7$\pm$22.8  &  0.02$\pm$0.56  &  12.94$\pm$3.6  &  24.71$\pm$0.07  &  23.78$\pm$0.07  &                       \\
20151005-1020.cand008  &  51.1$\pm$26.4  &  0.02$\pm$0.56  &  17.90$\pm$5.7  &  25.42$\pm$0.06  &  24.80$\pm$0.07  &                       \\
20151005-1020.cand009  &  48.2$\pm$24.9  &  0.02$\pm$0.56  &  6.52$\pm$0.8  &  25.10$\pm$0.04  &  24.83$\pm$0.18  &                       \\
20151005-1020.cand010  &  45.9$\pm$23.8  &  0.02$\pm$0.56  &  5.30$\pm$0.4  &  25.27$\pm$0.13  &  24.39$\pm$0.13  &                       \\
20151005-1020.cand011  &  46.6$\pm$24.2  &  0.02$\pm$0.56  &  5.49$\pm$0.3  &  24.60$\pm$0.40  &  23.76$\pm$0.06  &                       \\
20151005-1020.cand012  &  41.7$\pm$21.8  &  0.02$\pm$0.57  &  27.20$\pm$10.4  &  25.40$\pm$0.16  &  25.10$\pm$0.08  &                       \\
20151005-1020.cand013  &  40.4$\pm$21.2  &  0.03$\pm$0.57  &  21.14$\pm$7.4  &  25.16$\pm$0.14  &  24.46$\pm$0.09  &                       \\
20151005-1020.cand014  &  51.8$\pm$26.7  &  0.02$\pm$0.55  &  16.30$\pm$5.0  &  25.10$\pm$0.08  &  24.45$\pm$0.07  &                       \\
20151005-1020.cand015  &  35.6$\pm$18.8  &  0.03$\pm$0.58  &  20.59$\pm$7.5  &  24.38$\pm$0.02  &  23.76$\pm$0.24  &                       \\
20151005-1020.cand016  &  42.7$\pm$22.3  &  0.03$\pm$0.56  &  9.39$\pm$1.4  &  24.40$\pm$0.07  &  23.77$\pm$0.18  &                       \\
20151005-1020.cand017  &  46.8$\pm$24.2  &  0.02$\pm$0.56  &  8.56$\pm$0.7  &  24.79$\pm$0.10  &  23.81$\pm$0.05  &                       \\
20151005-1020.cand018  &  43.2$\pm$22.6  &  0.02$\pm$0.58  &  32.97$\pm$13.8  &  24.51$\pm$0.08  &  23.70$\pm$0.02  &                       \\
20151005-1020.cand019  &  46.5$\pm$24.1  &  0.02$\pm$0.56  &  9.72$\pm$1.6  &  24.44$\pm$0.08  &  23.83$\pm$0.05  &                       \\
20151005-1020.cand020  &  33.7$\pm$17.9  &  0.03$\pm$0.58  &  6.37$\pm$0.1  &  24.14$\pm$0.04  &  23.23$\pm$0.03  &                       \\
20151005-1020.cand021  &  35.1$\pm$18.6  &  0.03$\pm$0.58  &  16.29$\pm$5.3  &  24.96$\pm$0.13  &  24.44$\pm$0.13  &                       \\
20151005-1020.cand022  &  31.2$\pm$16.7  &  0.03$\pm$0.59  &  6.99$\pm$1.5  &  24.96$\pm$0.13  &  23.97$\pm$0.04  &                       \\
20151005-1020.cand023  &  46.8$\pm$24.2  &  0.02$\pm$0.56  &  5.04$\pm$0.1  &  24.73$\pm$0.27  &  23.66$\pm$0.09  &                       \\
20151005-1020.cand024  &  43.2$\pm$22.5  &  0.02$\pm$0.56  &  5.66$\pm$0.3  &  24.71$\pm$0.08  &  23.89$\pm$0.02  &                       \\
20151005-1020.cand025  &  35.3$\pm$18.7  &  0.03$\pm$0.58  &  10.54$\pm$3.0  &  24.56$\pm$0.11  &  23.93$\pm$0.05  &                       \\
20151005-1020.cand026  &  20.6$\pm$11.6  &  0.05$\pm$0.49  &  142.97$\pm$6.9  &  24.56$\pm$0.07  &  23.94$\pm$0.07  &                       \\
20151005-1020.cand027  &  46.6$\pm$24.2  &  0.02$\pm$0.56  &  9.84$\pm$2.8  &  23.59$\pm$0.07  &  23.09$\pm$0.06  &                       \\
20151005-1020.cand028  &  38.9$\pm$20.5  &  0.03$\pm$0.58  &  31.08$\pm$13.3  &  24.77$\pm$0.13  &  24.44$\pm$0.08  &                       \\
20151005-1020.cand029  &  39.6$\pm$20.8  &  0.03$\pm$0.57  &  20.84$\pm$7.3  &  23.29$\pm$0.07  &  22.70$\pm$0.02  &                       \\
20151005-1020.cand030  &  48.4$\pm$25.0  &  0.02$\pm$0.56  &  24.62$\pm$8.8  &  25.40$\pm$0.09  &  25.17$\pm$0.04  &                       \\
20151005-1020.cand031  &  44.2$\pm$23.0  &  0.02$\pm$0.56  &  6.89$\pm$0.5  &  25.28$\pm$0.18  &  25.05$\pm$0.39  &                       \\
20151005-1020.cand032  &  45.0$\pm$23.4  &  0.02$\pm$0.56  &  17.39$\pm$5.8  &  25.18$\pm$0.23  &  24.55$\pm$0.19  &                       \\
20151005-1020.cand033  &  41.8$\pm$22.6  &  0.02$\pm$0.66  &  53.20$\pm$39.6  &  23.93$\pm$0.69  &  22.63$\pm$0.07  &                       \\
20151005-1020.cand034  &  47.1$\pm$24.4  &  0.02$\pm$0.56  &  6.27$\pm$0.0  &  24.56$\pm$0.12  &  24.08$\pm$0.05  &                       \\
20151005-1020.cand035  &  31.9$\pm$17.0  &  0.03$\pm$0.59  &  11.34$\pm$2.9  &  24.61$\pm$0.04  &  24.05$\pm$0.11  &                       \\
20151005-1020.cand036  &  39.0$\pm$20.5  &  0.03$\pm$0.57  &  7.14$\pm$0.7  &  24.39$\pm$0.13  &  23.78$\pm$0.06  &                       \\
20151005-1020.cand037  &  46.2$\pm$24.0  &  0.02$\pm$0.57  &  26.40$\pm$9.9  &  25.19$\pm$0.15  &  25.05$\pm$0.14  &                       \\
20151005-1020.cand038  &  32.3$\pm$17.3  &  0.03$\pm$0.59  &  17.44$\pm$6.0  &  25.35$\pm$0.19  &  24.49$\pm$0.04  &                       \\
20151005-1020.cand039  &  45.7$\pm$23.7  &  0.02$\pm$0.56  &  11.25$\pm$2.7  &  25.36$\pm$0.17  &  25.00$\pm$0.61  &                       \\
20151005-1020.cand040  &  50.0$\pm$25.8  &  0.02$\pm$0.57  &  32.14$\pm$12.4  &  25.25$\pm$0.12  &  24.73$\pm$0.07  &                       \\
20151005-1020.cand041  &  44.4$\pm$23.1  &  0.02$\pm$0.56  &  7.92$\pm$0.0  &  24.76$\pm$0.06  &  24.40$\pm$0.09  &                       \\
20151005-1020.cand042  &  44.4$\pm$23.1  &  0.02$\pm$0.56  &  7.95$\pm$0.0  &  24.60$\pm$0.10  &  23.87$\pm$0.04  &                       \\
20151005-1020.cand043  &  38.1$\pm$20.0  &  0.03$\pm$0.57  &  10.95$\pm$1.6  &  23.92$\pm$0.16  &  22.98$\pm$0.07  &                       \\
20151005-1020.cand044  &  41.5$\pm$21.7  &  0.03$\pm$0.56  &  10.66$\pm$1.8  &  25.28$\pm$0.15  &  24.26$\pm$0.09  &                       \\
20151005-1020.cand045  &  35.4$\pm$18.7  &  0.03$\pm$0.58  &  9.98$\pm$1.6  &  24.44$\pm$0.02  &  23.84$\pm$0.09  &                       \\
20151005-1020.cand046  &  45.2$\pm$23.5  &  0.02$\pm$0.56  &  8.27$\pm$0.3  &  24.98$\pm$0.21  &  24.57$\pm$0.08  &                       \\
20151005-1020.cand047  &  55.8$\pm$28.6  &  0.02$\pm$0.55  &  22.77$\pm$7.4  &  25.02$\pm$0.13  &  24.36$\pm$0.03  &                       \\
20151005-1020.cand048  &  57.2$\pm$29.4  &  0.02$\pm$0.56  &  35.73$\pm$14.3  &  25.49$\pm$0.09  &  24.78$\pm$0.12  &                       \\
20151005-1020.cand049  &  35.4$\pm$18.7  &  0.03$\pm$0.58  &  10.74$\pm$2.1  &  24.19$\pm$0.04  &  23.60$\pm$0.09  &                       \\
20151005-1020.cand050  &  38.9$\pm$20.4  &  0.03$\pm$0.57  &  9.84$\pm$1.1  &  25.28$\pm$0.10  &  24.64$\pm$0.06  &                       \\
20151005-1020.cand051  &  35.5$\pm$18.8  &  0.03$\pm$0.57  &  8.68$\pm$0.9  &  25.13$\pm$0.13  &  24.23$\pm$0.05  &                       \\
20151005-1020.cand052  &  43.2$\pm$22.5  &  0.02$\pm$0.56  &  11.54$\pm$2.9  &  24.94$\pm$0.10  &  24.31$\pm$0.16  &                       \\
20151005-1020.cand053  &  37.3$\pm$19.6  &  0.03$\pm$0.57  &  7.18$\pm$0.2  &  24.58$\pm$0.10  &  24.01$\pm$0.07  &                       \\
20151005-1020.cand054  &  35.5$\pm$18.8  &  0.03$\pm$0.58  &  19.15$\pm$6.3  &  25.32$\pm$0.08  &  24.42$\pm$0.13  &                       \\
20151005-1020.cand055  &  45.7$\pm$23.8  &  0.02$\pm$0.57  &  32.60$\pm$13.2  &  24.51$\pm$0.11  &  23.73$\pm$0.17  &                       \\
20151005-1020.cand056  &  34.7$\pm$18.4  &  0.03$\pm$0.58  &  10.44$\pm$2.0  &  24.10$\pm$0.08  &  23.55$\pm$0.05  &                       \\
20151005-1020.cand057  &  40.1$\pm$21.0  &  0.02$\pm$0.58  &  25.02$\pm$9.3  &  25.38$\pm$0.16  &  25.36$\pm$0.27  &                       \\
20151005-1020.cand058  &  39.4$\pm$20.6  &  0.02$\pm$0.58  &  29.40$\pm$9.0  &  25.32$\pm$0.11  &  24.58$\pm$0.11  &                       \\
20151005-1020.cand059  &  42.7$\pm$22.2  &  0.02$\pm$0.56  &  14.36$\pm$0.4  &  23.51$\pm$0.03  &  22.75$\pm$0.02  &                       \\
20151005-1020.cand060  &  40.8$\pm$21.3  &  0.02$\pm$0.57  &  14.73$\pm$1.3  &  23.91$\pm$0.04  &  23.07$\pm$0.01  &                       \\
20151005-1020.cand061  &  38.7$\pm$20.3  &  0.03$\pm$0.57  &  19.13$\pm$3.4  &  24.95$\pm$0.04  &  24.26$\pm$0.08  &                       \\
20151005-1020.cand062  &  47.0$\pm$24.3  &  0.02$\pm$0.56  &  15.62$\pm$0.0  &  24.99$\pm$0.07  &  24.31$\pm$0.08  &                       \\
20151005-1020.cand063  &  45.2$\pm$23.4  &  0.02$\pm$0.56  &  17.29$\pm$1.8  &  24.15$\pm$0.08  &  23.18$\pm$0.04  &                       \\
20151005-1020.cand064  &  33.9$\pm$18.0  &  0.03$\pm$0.59  &  29.57$\pm$8.8  &  25.21$\pm$0.13  &  24.67$\pm$0.08  &                       \\
20151005-1020.cand065  &  55.1$\pm$28.2  &  0.02$\pm$0.55  &  15.03$\pm$0.0  &  25.40$\pm$0.11  &  24.78$\pm$0.03  &                       \\
20151005-1020.cand066  &  47.0$\pm$24.3  &  0.02$\pm$0.56  &  14.60$\pm$0.0  &  25.22$\pm$0.06  &  24.69$\pm$0.08  &                       \\
20151005-1020.cand067  &  41.4$\pm$21.6  &  0.02$\pm$0.58  &  29.07$\pm$9.8  &  24.96$\pm$0.12  &  24.44$\pm$0.07  &                       \\
20151005-1020.cand068  &  41.1$\pm$21.6  &  0.03$\pm$0.60  &  42.99$\pm$21.5  &  25.34$\pm$0.13  &  24.83$\pm$0.14  &                       \\
20151005-1020.cand069  &  43.6$\pm$22.7  &  0.02$\pm$0.58  &  33.98$\pm$13.0  &  25.44$\pm$0.11  &  25.06$\pm$0.05  &                       \\
20151005-1020.cand070  &  38.8$\pm$20.3  &  0.02$\pm$0.57  &  15.74$\pm$0.5  &  25.36$\pm$0.16  &  24.99$\pm$0.14  &                       \\
20151005-1020.cand071  &  38.6$\pm$20.2  &  0.03$\pm$0.57  &  16.08$\pm$1.6  &  24.58$\pm$0.08  &  23.80$\pm$0.13  &                       \\
20151005-1020.cand072  &  34.8$\pm$18.4  &  0.03$\pm$0.58  &  14.74$\pm$0.4  &  24.08$\pm$0.20  &  23.31$\pm$0.07  &                       \\
20151005-1020.cand073  &  43.3$\pm$22.5  &  0.02$\pm$0.57  &  21.56$\pm$4.7  &  25.06$\pm$0.12  &  24.41$\pm$0.04  &                       \\
20151005-1020.cand074  &  70.0$\pm$35.5  &  0.01$\pm$0.54  &  15.11$\pm$0.3  &  25.17$\pm$0.11  &  24.48$\pm$0.09  &                       \\
20151005-1020.cand075  &  48.1$\pm$24.9  &  0.02$\pm$0.58  &  39.92$\pm$14.7  &  24.63$\pm$0.12  &  23.79$\pm$0.06  &                       \\
20151005-1020.cand076  &  39.1$\pm$20.4  &  0.03$\pm$0.57  &  14.27$\pm$0.1  &  25.36$\pm$0.15  &  25.02$\pm$0.15  &                       \\
20151005-1020.cand077  &  43.1$\pm$22.3  &  0.02$\pm$0.56  &  14.69$\pm$0.1  &  25.61$\pm$0.16  &  24.93$\pm$0.10  &                       \\
20151005-1020.cand078  &  30.2$\pm$16.2  &  0.03$\pm$0.59  &  14.98$\pm$0.3  &  25.65$\pm$0.22  &  25.02$\pm$0.09  &                       \\
20151005-1020.cand079  &  31.7$\pm$16.9  &  0.03$\pm$0.59  &  18.37$\pm$2.3  &  25.06$\pm$0.10  &  24.12$\pm$0.04  &                       \\
20151005-1020.cand080  &  33.8$\pm$17.9  &  0.03$\pm$0.59  &  23.50$\pm$6.8  &  23.44$\pm$0.02  &  22.92$\pm$0.03  &                       \\
20151005-1020.cand081  &  39.5$\pm$20.6  &  0.03$\pm$0.57  &  17.19$\pm$1.9  &  23.43$\pm$0.03  &  22.84$\pm$0.03  &                       \\
20151005-1020.cand082  &  41.5$\pm$21.6  &  0.02$\pm$0.57  &  16.86$\pm$1.6  &  24.56$\pm$0.02  &  23.83$\pm$0.03  &                       \\
20151005-1020.cand083  &  38.2$\pm$20.0  &  0.03$\pm$0.57  &  18.99$\pm$1.9  &  25.28$\pm$0.13  &  24.81$\pm$0.12  &                       \\
20151005-1020.cand084  &  39.7$\pm$20.7  &  0.02$\pm$0.57  &  27.22$\pm$6.4  &  24.31$\pm$0.05  &  23.68$\pm$0.03  &                       \\
20151005-1020.cand085  &  20.7$\pm$11.5  &  0.05$\pm$0.50  &  156.85$\pm$2.0  &  25.22$\pm$0.11  &  24.50$\pm$0.22  &                       \\
20151005-1020.cand086  &  39.9$\pm$20.8  &  0.02$\pm$0.57  &  17.57$\pm$0.9  &  24.56$\pm$0.05  &  23.94$\pm$0.09  &                       \\
20151005-1020.cand087  &  48.4$\pm$25.0  &  0.02$\pm$0.56  &  18.40$\pm$1.4  &  24.83$\pm$0.27  &  24.46$\pm$0.14  &                       \\
20151005-1020.cand088  &  39.6$\pm$20.7  &  0.02$\pm$0.57  &  20.61$\pm$2.2  &  23.35$\pm$0.04  &  22.46$\pm$0.04  &                       \\
20151005-1020.cand089  &  45.3$\pm$23.4  &  0.02$\pm$0.56  &  16.99$\pm$1.2  &  25.56$\pm$0.15  &  24.89$\pm$0.08  &                       \\
20151005-1020.cand090  &  52.4$\pm$26.9  &  0.02$\pm$0.56  &  17.49$\pm$1.2  &  24.97$\pm$0.14  &  24.60$\pm$0.10  &                       \\
20151005-1020.cand091  &  46.9$\pm$24.2  &  0.02$\pm$0.57  &  23.63$\pm$5.9  &  24.99$\pm$0.10  &  24.48$\pm$0.06  &                       \\
20151005-1020.cand092  &  40.1$\pm$20.9  &  0.03$\pm$0.58  &  26.83$\pm$8.3  &  25.25$\pm$0.07  &  24.67$\pm$0.10  &                       \\
20151005-1020.cand093  &  41.1$\pm$21.4  &  0.02$\pm$0.58  &  27.66$\pm$8.7  &  24.81$\pm$0.14  &  24.11$\pm$0.11  &                       \\
20151005-1020.cand094  &  39.9$\pm$20.8  &  0.03$\pm$0.59  &  10.95$\pm$4.5  &  24.78$\pm$0.17  &  24.10$\pm$0.07  &                       \\

\hline
\end{longtable}


\end{document}